\documentclass[12pt]{article}
\usepackage{amsmath,amssymb,bm,epsfig,epsf,graphicx,float,url,multirow,hyperref}
\usepackage{dsfont}
\usepackage{caption}
\usepackage{subcaption}
\input epsf.sty
\newcommand{\Tr}{\mathop{\rm Tr}\nolimits}

\def\bra#1{\langle #1 |}
\def\bbraa#1{\langle\!\langle \hspace{0.5pt}  #1 \hspace{0.5pt}|\hspace{-1pt}|}
\def\ket#1{|#1 \rangle}
\def\kett#1{|#1 \rangle\!\rangle}
\def\kkett#1{|\hspace{-1pt}|\hspace{0.5pt} #1 \hspace{0.5pt}\rangle\!\rangle}
\def\aver#1{\left\langle\, #1 \,\right\rangle}

\def\half{\frac{1}{2}}

\let\eps = \varepsilon
\def \be {\begin{equation}}
\def \ee {\end{equation}}
\def \bea {\begin{eqnarray}}
\def \eea {\end{eqnarray}}
\def \bdm {\begin{displaymath}}
\def \edm {\end{displaymath}}
\def \ot{\otimes}

\def \vv {{\cal V}}

\def\VVV#1#2#3{\langle #1,#2,#3\rangle}
\def\VV#1#2{\langle #1,#2\rangle}

\def \la {\langle}
\def \ra {\rangle}

\begin{document}
\vskip 2.1cm

\centerline{\large \bf Ising model conformal boundary conditions from open string field theory}
\vspace*{8.0ex}

\centerline{\large \rm Mat\v{e}j Kudrna\footnote{Email: {\tt kudrna at fzu.cz}},
Miroslav Rap\v{c}\'{a}k\footnote{Email: {\tt miroslav.rapcak at gmail.com}},
Martin Schnabl\footnote{Email: {\tt schnabl.martin at gmail.com}}}

\vspace*{8.0ex}

\centerline{\large \it Institute of Physics AS CR, Na Slovance 2, Prague 8, Czech Republic}
\vspace*{2.0ex}

\vspace*{6.0ex}

\centerline{\bf Abstract}
\bigskip

 Given a consistent choice of conformally invariant boundary conditions in a two dimensional conformal field theory, one can construct new consistent boundary conditions by deforming with a relevant boundary operator and flowing to the infrared, or by a marginal deformation. Open string field theory provides a very universal tool to discover and study such new boundary theories. Surprisingly, it also allows one to go in the reverse direction and to uncover solutions with higher boundary entropy. We will illustrate our results on the well studied example of Ising model.

 \vfill \eject

\baselineskip=16pt

\tableofcontents

\setcounter{footnote}{0}
\section{Introduction and summary}

Two-dimensional conformal field theory (CFT) is a rich subject with a long history. It has proved important for applications ranging from the fundamental description of string dynamics to condensed matter systems. For a number of important applications (e.g. study of open strings on D-branes, surface critical behavior in lattice models, or impurities in condensed matter models) the theory should be formulated on spaces with boundaries, and a question arises what are the possible boundary conditions allowed by the conformal invariance.\footnote{For a general introduction to the boundary conformal field theory we refer to the reviews \cite{CardyReview, PZ, RunkelPhD, GaberdielReview, SchmidtColinetPhD} and to the very recent book \cite{RS}.} To address this question, it is useful to represent the boundary conditions via boundary states, in the usual setup, where the radial coordinate on the disk is reinterpreted as time coordinate.


For a given bulk theory, the allowed conformal boundary states are required to obey a number of necessary conditions. The most basic requirement is that the two-dimensional energy and momentum do not flow in or out across the boundary (at any given point along it), which is expressed through the so called gluing condition:
\be
\left(L_n - \bar L_{-n} \right) \kkett{B} = 0.
\label{gluing}
\ee
This equation was solved in full generality long time ago by Ishibashi \cite{Ishibashi}. The solutions are called the Ishibashi states and are in one-to-one correspondence with the set of spinless bulk primaries $V^\alpha$
\bea
\kett{V_\alpha} &=& \sum_{IJ} M^{IJ}(h_\alpha) L_{-I}\bar L_{-J} \ket{V_\alpha} \\
&=& \sum_n \ket{n,\alpha} \otimes \overline{\ket{n,\alpha}} \\
&=& \Big[ 1+\frac1{2h_\alpha}L_{-1}\bar L_{-1} + \cdots \Big] \ket{V_\alpha}.
\eea
In the first line the multi-indices $I,J$ label the non-degenerate descendants in the conformal family of $V_\alpha$,  and  $M^{IJ}(h_\alpha)$ is defined
as the inverse of the real symmetric matrix $\bra{V^\alpha}L_{I}L_{-J}\ket{V_\alpha}$. For the multi-index $I=\{i_1,i_2,\dots i_n\}$, where $0 \le i_1 \le i_2 \le \cdots \le i_n$ we define $L_{I} = L_{i_1} L_{i_2} \ldots L_{i_n}$ and $L_{-I} = L_{-i_n} L_{-i_{n-1}} \ldots L_{-i_1}$. Second line expresses the Ishibashi state as a sum over a complete orthonormal basis of states in the Verma module over the chiral part of the primary $V_\alpha$. As an illustration, the last line shows the first two terms in the case of nonzero conformal weight $h_\alpha$ of the primary $V_\alpha$ .

Second condition, which consistent boundary states must satisfy, comes from considering cylindrical surface. Interpreting the linear dimension as time, the path integral on the surface can be interpreted as a matrix element between two boundary states $\kkett{a}$ and $\kkett{b}$. Interpreting the angular direction as time, the same path integral can be interpreted as a trace over the Hilbert space of the CFT with the two boundary conditions $a$ and $b$:
\be
\bbraa{a} {\tilde q}^{\frac{1}{2}(L_0+\bar{L}_0 -\frac{c}{12})} \kkett{b} = \Tr_{{\cal H}_{ab}^{\mathrm{open}}} \left( q^{L_0-\frac{c}{24}} \right),
\label{cardy's condition}
\ee
where
\be
q= e^{2\pi i \tau}, \qquad\qquad \tilde q =e^{-2\pi i /\tau}
\ee
and $\tau=R/L$ being the modular parameter of the cylinder expressed in terms of its radius and length.

Whatever boundary conditions, for theories with discrete spectrum, the power series in $q=e^{2\pi i \tau}$ on the right hand side must have non-negative integer coefficients. This then places strong constraints on the possible form of the boundary states $\kkett{a}$ and $\kkett{b}$ on the left hand side, which thus cannot be arbitrary linear combinations of Ishibashi states.
In the case of minimal models with diagonal modular invariant partition functions, this problem was solved long ago by Cardy \cite{cardy-boundary operators}.\footnote{For the discussion in more general theories see \cite{PSS,BPPZ} and the reviews \cite{PZ,RunkelPhD}.} The right hand side can be expressed as a sum over Virasoro characters
\be
\sum_i n_{ab}^i \chi_i(q).
\ee
To match with the left hand side, one can perform modular transformation $\tau \to -1/\tau$ upon which the characters transform as $\chi_i(q) = S_i^{\; j} \chi_j(\tilde q)$.
With the help of the Verlinde formula \cite{Verlinde}
\be\label{Verlinde}
\sum_k S_k^{\;p} N_{ij}^{\;\;k} = \frac{S_i^{\;p}S_j^{\;p}}{S_0^{\;p}}
\ee
Cardy found a very elegant solution
\be
\kkett{B_i} = \sum_j \frac{S_{i}^{\; j}}{\sqrt{S_{0}^{\; j}}} \, \kett{j}
\label{Cardy_sol}
\ee
for a set of fundamental boundary states labeled by the irreducible representations of the Virasoro algebra. More general boundary states can be constructed by taking linear combinations of such states with non-negative integer coefficients. These form the complete solution for a class of rational CFT's with diagonal modular invariant partition function.

The possible boundary operators which can be inserted between segments with boundary conditions $i$ and $j$ in the minimal models (i.e. the spectrum of open strings stretched between D-branes $\kkett{i}$ and $\kkett{j}$) are those with labels which appear on the right hand side of the fusion algebra
\be
\phi_i \times \phi_j = \sum_k N_{ij}^{\;\;\,k} \, \phi_k
\label{fusion}
\ee
and their multiplicity is given precisely by the fusion coefficients $N_{ij}^{\;\;\,k}$.


The allowed boundary conditions and the associated boundary and bulk-boundary OPE coefficients are further constrained by four additional sewing constraints \cite{Lewellen} (three of them for the disk, one for the cylinder) which guarantee consistency of conformal field theory on an arbitrary Riemann surface with boundaries. These conditions are typically harder to solve, but often more powerful than just the Cardy condition itself.

\vskip20pt

Open string field theory \cite{Witten}\footnote{Classical reviews are \cite{Thorn, TZ, SenRev}, while the recent developments are reviewed in \cite{FK, lightning, OkawaRev}.} was for a long time considered to be a dynamical theory of open strings, but in late nineties it was understood, that one should better think of it as theory describing D-branes. The fields living on a given D-brane can take various configurations, and the combined object can represent a completely different D-brane. Since classification of D-branes, or equivalently boundary states, in a given string theory background is a notoriously hard problem, a natural question arises if one can use string field theory to search for them systematically. So far, there is a numerical evidence for a few D-branes appearing in this way: Dirichlet D-brane from Neumann brane in $c=1$ and $c=2$ free boson models \cite{MSZ,Moeller,KMS} and analogous solutions on the $SU(2)$ group manifold \cite{Michishita}. There are of course numerical solutions for marginal deformations \cite{SZmarg,Kurs,CZJPmarg}.  Analytically, only a handful of fully regular solutions have been constructed: the tachyon vacuum \cite{Schnabl, Okawa} and the marginal deformation solutions \cite{marg,KORZ,KO}. There were also interesting proposals for solutions describing the endpoint of a relevant flow  \cite{BMT} and for the multiple D-brane configurations \cite{Murata1, Murata2}.\footnote{These proposals apparently subtly violate the equation of motion when contracted with Fock states and we hope that better solutions will eventually be found.} All of the above results have been obtained first in the bosonic string field theory, but many of them have been generalized to the superstring theory in the subsequent works.

One of the major questions of string field theory is how "distant" boundary state can a classical solution describe. In the absence of adequate analytic techniques, we attempt at a systematic numerical exploration of the space of solutions of level truncated equations of motion and match them with the known boundary states. For definiteness we focus our attention on solutions describing change of boundary conditions in $c=\half$ Ising model and $c=1$ $(\mathrm{Ising})^2$ conformal field theories. Our general strategy for studying changes in boundary conditions in such models is to formulate an OSFT on the background described in the matter sector by $\mathrm{BCFT}_c \otimes \mathrm{BCFT}_{26-c}$ boundary conformal field theory, where the first factor is the CFT of interest with given boundary conditions, whereas the second one serves to provide a consistent string theory background without the conformal anomaly. The solutions which we look for numerically can take the most general form in the first BCFT sector, but are restricted to live in the universal Verma module of the identity in the second sector.

A recent key element which allows us to identify easily various solutions and thus makes the whole program feasible is the practical construction of the corresponding boundary state \cite{KMS} built upon Ellwood's interpretation  \cite{Ellwood} of certain gauge invariant observables. An alternative more geometric construction for the boundary state has been put forward also in \cite{KOZ} but it is not clear how to apply it to solutions known only numerically. It would be very interesting to explore the relationship between these two constructions.

Ideally, we would like to be able to find all classical solutions in a given OSFT and construct the corresponding boundary states. The complexity of the truncated system of equations of motion grows quite rapidly with the level. For example for the Ising model at level 24 on the so called $\mathds{1}$-brane the system comprises of 82309 coupled quadratic equations which could have a priori up to $2^{82309}$ complex solutions, which is clearly impossible to explicitly describe by any means. Our strategy is to systematically explore all solutions at level 4 where we have 13 equations with $8091$ complex or real solutions (curiously less than $2^{13}=8192$ which is a number one would expect for such a system if the coefficients were generic) and see how these get improved when used as starting points for Newton's iteration method at higher levels. If we allow complex starting points, with the hope of approaching real solution as the level is increased, we do indeed find number of interesting solutions. Aside of the perturbative and tachyon vacua, we surprisingly find on the $\mathds{1}$-brane a {\it real} solution describing the $\sigma$-brane which has higher energy. It is the first such example discovered in string field theory and it shows its power to go "against" the RG flow. We found other interesting solutions which might be interpreted as the $\eps$-brane and perhaps even some integer combinations of the fundamental branes, although in these cases the physical invariants are harder to reliably extrapolate to infinite level and the interpretation is less straightforward.

On the $\sigma$-brane the spectrum of boundary operators is richer, so in addition to the solutions in the Verma module of the identity, which look exactly the same as on the $\mathds{1}$-brane, there are solutions which turn on fields in the Verma module of the $\eps$ boundary field with dimension $1/2$. The simplest solutions, analogous in some respects to lump solutions on the circle, describe the $\mathds{1}$-brane and the $\eps$-brane. These solutions with exactly the same coefficients at every level can be found in OSFT built on $(\mathrm{Ising})^2$ BCFT, where they can be interpreted via the orbifold correspondence as bulk D0-branes decaying into fractional D1-branes.

The existence of the solutions in the Verma module of the identity, which can have one interpretation on one D-brane and another interpretation on a different D-brane, leads to a very interesting corollary
\be\label{boundary states relation}
\frac{B_x^\beta B_y^\beta}{B_0^\beta} = \sum_z N_{xy}^{\;\;\;z} B_z^\beta
\ee
for the coefficients $B_w^\beta$ of the boundary states $\ket{B_w}$. The index $\beta$ labels the spinless closed string primary states $V^\beta$ and the corresponding Ishibashi states.
The coefficients $N_{xy}^{\;\;\;z}$ on the right hand side ought to be positive integers.

This relation is derived under the assumption that the boundary state $B_x$ can be found as a solution of OSFT formulated around D-brane described by the boundary state $B_0$ and that this solution can be reinterpreted on a D-brane with the boundary state $B_y$ as a new linear integer combination of boundary states $B_z$ in the same theory. The factor $\frac{B_y^\beta}{B_0^\beta}$  on the left hand side of (\ref{boundary states relation}) comes from a relative normalization and the fact that the prescription given in \cite{KMS} is linear in the string field.
To be able to reinterpret the solution found on $B_0$ as a solution on $B_y$ it is necessary, that the solution on $B_0$ switches on only the boundary operators which are present on $B_y$. For instance, if all D-branes could be found on the $\mathds{1}$-brane of some minimal model theory, then it is guaranteed that they can be reinterpreted on any other D-brane of the same theory, since all of them include the Verma module of the identity. In such a case the formula (\ref{boundary states relation}) would be valid for all $x$ and $y$.

A remarkable consequence of (\ref{boundary states relation})---if valid for all $x$ and $y$---is, in particular, that the set of tensions of the D-branes normalized by the tension of the $\mathds{1}$-brane would have both an additive and multiplicative structure, forming thus a commutative semiring. In the case of minimal models, it is a simple consequence of the Verlinde formula (\ref{Verlinde}) and the explicit Cardy solution
(\ref{Cardy_sol}), but surprisingly it remains true also for the non-diagonal Potts model \cite{AOS,Fuchs:1998qn,BPPZ}.\footnote{We have verified the formula (\ref{boundary states relation}) for all eight boundary states of the Potts model. The only tricky point was to choose the right bases when some spinless primaries have degenerate representation.}  While the appearance of Verlinde formula in CFT is somewhat enigmatic, in the context of OSFT the formula (\ref{boundary states relation}) is a simple consequence of keeping track of the overall normalization when reinterpreting the same solution on different branes.

\vskip20pt

The rest of this letter is organized as follows: After the exposition of our general strategy we illustrate it on various Ising model related examples. We start by reviewing the BCFT for the three conformal boundary conditions called $\mathds{1}, \eps, \sigma$ and outline the details of our computation. In section \ref{sec:sigma} we study the tachyon condensation on the $\sigma$-brane triggered by the relevant boundary operator $\eps$. This case is more-or-less analogous to the ordinary lower dimensional D-brane formation in the free boson CFT. In section \ref{sec:identity} we start with the $\mathds{1}$- or $\eps$-brane, and find rather surprisingly that there exists a well behaved solution with higher energy which should be interpreted as the $\sigma$-brane. Finally, in section \ref{sec:double} we apply our methods to the tensor product of two Ising models, which is useful for studying conformal defects in the Ising model, but at the same time, is dual to the free boson on the $S^1/Z_2$ orbifold. In appendix \ref{ap: solutions} we present further less unambiguous numerical solutions for the Ising model and in appendix \ref{ap: double_ising} we review some aspects of the double Ising model and its D-branes.

\section{Preliminary notions}
\setcounter{equation}{0}

\subsection{Ising model review}
In this section, we review few basic facts about the Ising model that will become necessary ingredients for the following construction of boundary states. Detailed discussion can be found in most of the CFT textbooks, see e.g. \cite{francesco,ginsparg}. The Ising model is an example of the simplest but yet nontrivial minimal CFT model with central charge $c=\frac{1}{2}$. The operator spectrum consists of three Verma modules corresponding to the three primary operators usually denoted as $\mathds{1}$, $\eps$, and $\sigma$ with conformal weights (0,0), $\left (\frac{1}{2},\frac{1}{2}\right )$, and  $\left (\frac{1}{16},\frac{1}{16}\right )$ and fusion rules of the form
\begin{equation}
\eps \times \eps = \mathds{1},\qquad
\sigma \times \sigma = \mathds{1} + \eps, \qquad
\sigma \times \eps = \sigma.
\label{fusion ising}
\end{equation}

Introducing a boundary to the CFT one has to impose a consistent boundary condition, which can be encoded as a boundary state in the radial quantization scheme. Boundary states can be looked for by solving a set of consistency conditions already mentioned in the introduction. For each spinless primary field $V_{\alpha}$ we can find level-by-level the corresponding Ishibashi state satisfying gluing conditions (\ref{gluing}). The consistent boundary states are given by their specific linear combinations. In the case of the Ising model we have three Ishibashi states and the possible boundary states are of the form
\begin{equation}
\kkett{B_a} = B_a^{\mathds{1}}\kett{\mathds{1}}+B_a^{\eps}\kett{\eps}+B_a^{\sigma}\kett{\sigma}.
\label{boundary}
\end{equation}
The coefficients $B_i^\alpha$ have been determined by solving Cardy's consistency condition relating two different ways of computing partition function of the model on the cylinder (\ref{cardy's condition}). Cardy's solution (\ref{Cardy_sol}) is given in terms of the matrix elements of the modular S-matrix which is well known. In the case of the Ising model we find
\begin{eqnarray}
\kkett{\mathds{1}} &=&\frac{1}{\sqrt{2}}\kett{\mathds{1}} +\frac{1}{\sqrt{2}}\kett{\eps}+\frac{1}{\sqrt[4]{2}}\kett{\sigma}\nonumber\\
\kkett{\eps} &=&\frac{1}{\sqrt{2}}\kett{\mathds{1}}+\frac{1}{\sqrt{2}}\kett{\eps}-\frac{1}{\sqrt[4]{2}}\kett{\sigma}\nonumber\\
\kkett{\sigma} &=&\kett{\mathds{1}}-\kett{\eps}.
\label{boundary states}
\end{eqnarray}
Moreover, linear combinations of the boundary states with integer coefficients still satisfy Cardy's condition and thus describe consistent boundary condition. The first two boundary states differing by a relative sign in frond of the $\kett{\sigma}$ Ishibashi state can be interpreted as fixed boundary conditions in the lattice picture of the model (imposing $+$ or $-$ values of the spins along the boundary),  whereas in the third case we have free boundary condition. Inspired by string theory language we will call the first two boundary states simply as $\mathds{1}$-brane and $\eps$-brane and the last one $\sigma$-brane.

CFT with a boundary admits a spectrum of boundary operators \cite{cardy-boundary operators} which either preserve or change boundary conditions. In string theory these would be interpreted as vertex operators for open strings ending on a given D-brane, or stretching between two different D-branes labeled as $a$ and $b$. For the minimal models the boundary spectrum is encoded in the fusion rules (\ref{fusion}). The boundary primaries which change the boundary condition from $a$ to $b$ carry the labels of operators appearing on the right hand side of the fusion rules and the coefficients $N_{ab}^{\;\;c}$ tell us the multiplicity of the $c$-labeled operators in the spectrum. For the special case $a=b$, one finds from the integer $N_{aa}^{\;\;c}$ the
representations and multiplicities of boundary primaries which preserve the boundary condition $a$. From the Ising model fusion rules (\ref{fusion ising}) we see that $\mathds{1}$- and $\eps$-brane admit the identity operator as the only boundary primary operator, while the $\sigma$-brane admits also the boundary primary called $\eps$ with boundary scaling dimension $h_{\eps}=\frac{1}{2}$. 

D-branes in string theory are characterized not only by the spectra of their fluctuations, but also by their couplings to the closed strings. Perhaps the most prominent quantity is their energy or tension, which in the static case is given simply, up to an overall normalization, by the overlap of the boundary state with $\ket{0}$, or equivalently by the coefficient of the $\kett{\mathds{1}}$ Ishibashi state. This coefficient in the BCFT literature is called the universal noninteger 'ground state degeneracy' \cite{AL}, or simply the $g$-function. In table \ref{tab:Ising-branes} we summarize these results for the Ising model.
\begin{table}[th]
\centering
\begin{equation}
\begin{array}{|l|c|l|}\hline
& & \\[-1.5ex]
\mbox{D-brane}   & \mbox{Energy} & \mbox{Boundary spectrum}\\[0.5ex]\hline
 & & \\[-2ex]
 \kkett{\mathds{1}} & \frac{1}{\sqrt{2}} & \mathds{1}\\[0.5ex]
 \kkett{\eps} & \frac{1}{\sqrt{2}} & \mathds{1}\\[0.5ex]
 \kkett{\sigma} & 1 & \mathds{1}, \, \eps\\[0.5ex]
 \hline
\end{array}\nonumber
\end{equation}
\caption{List of Ising model D-branes with their tensions and spectra of boundary primary operators.}
\label{tab:Ising-branes}
\end{table}

Our goal in this work is to recover the boundary states (\ref{boundary states}) by solving string field theory equations of motion. Due to Sen's second conjecture, solutions to the equations of motion correspond to different BCFT backgrounds characterized by some boundary condition and its associated boundary state. To find such solutions in string field theory, we have to be equipped by the boundary three-point functions, as well as by the bulk-boundary structure constants in a particular BCFT background. Let us for definiteness work on the upper-half plane.

Recall that the OPE for bulk primaries present in the Ising model is of the form
 \begin{eqnarray}\nonumber
\eps(z,\bar z)\eps(w,\bar w)&\sim &\frac{1}{|z-w|^2},\\ \nonumber
\sigma(z,\bar z)\sigma(w,\bar w)&\sim &\frac{1}{|z-w|^\frac{1}{4}}+\frac{1}{2}|z-w|^{\frac{3}{4}}\eps(w,\bar w),\\
\eps(z,\bar z)\sigma(w,\bar w)&\sim &\frac{1}{2}\frac{1}{|z-w|}\sigma(w,\bar w),
\end{eqnarray}
where the normalization has been chosen so that $\langle V^{\alpha}|V_{\beta}\rangle =\delta _{\beta}^{\alpha}$ and $\aver{1}=1$ in the bulk. 
On the $\sigma$-brane, one has one  more nontrivial OPE for the boundary operator $\eps$ 
\begin{equation}
\eps (u)\eps (v)\sim \frac{1}{u-v}.
\label{three-pointOPE}
\end{equation}
In general, the boundary OPEs determine the boundary three-point function 
\begin{equation}
\langle \phi _i (u)\phi _j(v)\phi _k (w)\rangle ^a _{\text{UHP}} = \frac{C^a_{ijk}\langle \mathds{1}\rangle ^a_{UHP}}{(u-v)^{h_i+h_j-h_k}(u-w)^{h_i+h_k-h_j}(v-w)^{h_j+h_k-h_i}}
\label{three-point}
\end{equation}
for $u>v>w$ real. To simplify our notation, let us consider only the fields that do not change a given boundary condition $a$. Then the boundary structure constants $C^a_{ijk}$  for the Ising model are all equal to $1$, except for the $\sigma$-brane with odd number of $\eps$ boundary insertions, when they vanish.

The last thing we have to address are the bulk-boundary correlation functions. For a boundary fields $\phi _i$ with boundary scaling dimensions $h_i$ and a bulk field $V_\alpha$ with bulk scaling dimension $\Delta _{\alpha}= h_{\alpha} + \bar{h}_{\alpha}$ we have the following form of the bulk-boundary OPE
\begin{eqnarray}
V_{\alpha}(x+iy)\sim \sum _i B^{a}_{\alpha i}(2y)^{h_i-\Delta _{\alpha}}\phi _i(x),
\end{eqnarray}
where $B^{a}_{\alpha i}$ are the bulk-boundary structure constants \cite{cardy-bulk-boundary}.\footnote{A small deficiency of our notation is the unnatural looking relation $B^{a}_{\alpha \mathds{1}} = B_a^\alpha$.} For the three types of the allowed boundary conditions we have the following structure constants
\begin{eqnarray}\nonumber
&B^\mathds{1}_{\mathds{1}\mathds{1}}=B^\mathds{1}_{\eps\mathds{1}}=1, \quad B^\mathds{1}_{\sigma\mathds{1}}=\sqrt[4]{2},\\ \nonumber &B^\eps_{\mathds{1}\mathds{1}}=B^\eps_{\eps\mathds{1}}=1, \quad B^\eps_{\sigma\mathds{1}}=-\sqrt[4]{2}, \\
&B^\sigma_{\mathds{1}\mathds{1}}=1,\quad B^\sigma_{\eps\mathds{1}}=-1, \quad B^\sigma_{\sigma\mathds{1}}=B^\sigma_{\mathds{1}\eps}=B^\sigma_{\eps\eps}=0 ,\quad B^\sigma_{\sigma\eps}=\frac{1}{\sqrt[4]{2}}.
\end{eqnarray}
With the knowledge of the bulk-boundary structure constants the bulk-boundary correlator can be easily computed using the formula
\begin{equation}
\langle V_\alpha(x+iy) \phi _i (0)\rangle_{\text{UHP}} ^a = \frac{B^{a}_{\alpha i}\langle \mathds{1}\rangle ^a_{UHP}}{(2y)^{\Delta_\alpha-h_i}(x^2+y^2)^{h_i}}.
\label{bulk-boundary}
\end{equation}

\subsection{Computation setup}

Now, being equipped with the above correlators we can follow the procedure proposed in \cite{KMS} to obtain the coefficients $B_a^\alpha$ determining the boundary state (\ref{boundary}) by solving equations of motion in open string field theory. Let us take the matter sector of the OSFT to be 
of the form BCFT$^m$=BCFT$_I\otimes$BCFT$_R$, where BCFT$_I$ is the Hilbert space of the Ising model while  BCFT$_R$ is the Hilbert space of an additional sector with central charge $c=\frac{51}{2}$ to ensure that the total central charge in the matter sector is $c=26$. 
This is necessary for the consistency of OSFT. Upon adding the ghost sector BCFT$^{gh}$ with central charge $c=-26$ the BRST charge $Q$ becomes nilpotent and the total central charge vanishes.

Starting from an arbitrary  BCFT background with given boundary condition we can look for solutions to OSFT equations of motion $Q\Psi + \Psi*\Psi=0$ to find other possible boundary conditions. If $\Psi$ is a solution and $\Psi _{TV}$ is the tachyon vacuum solution then the coefficients  $B_a^\alpha$
are equal to the Ellwood invariants
\begin{equation}
n_{\Psi}^\alpha =2\pi i \langle E [\vv ^{\alpha}]\ket{\Psi - \Psi _{TV}}=2\pi i \langle I|\vv ^{\alpha}(i,-i)\ket{\Psi - \Psi _{TV}},
\label{coef}
\end{equation}
where $\vv ^{\alpha}=c\bar{c}V^{\alpha}\otimes w^{\alpha}$ for a bulk primary $V^{\alpha}$  with conformal weight $(h_{\alpha},\bar{h}_{\alpha})$ in the sector BCFT$_I$ and $w^\alpha$ is an auxiliary bulk primary with conformal weights $(1-h_{\alpha},1-\bar{h}_{\alpha})$ with identity disk one-point function in BCFT$_R$, i.e.  $\langle w^{\alpha}(0,0)\rangle _{\text{disk}}^R = 1$.

Imposing the Siegel-gauge condition $b_0 \Psi =0$ and the very useful $SU(1,1)$ singlet condition \cite{Zwiebach:trimming, Gaiotto}, the string field can be conveniently written in full generality as
\begin{equation}
\Psi - \Psi _{TV}=\sum _j \sum _{I,J,K}a_{IJK}^jL^{I}_{-I}\ket{\phi ^j} \otimes L^{R}_{-J}\ket0 \otimes L'^{gh}_{-K}c_1\ket{0},
\label{solution}
\end{equation}
where $j$ runs over all boundary primaries present in the given BCFT background, and $I,J,K$ are the multiindices labeling the descendants. Using conservation laws derived in the appendix of \cite{KMS} we can always end up with a linear combination of the following overlaps
\begin{equation}
\langle E[\vv ^{\alpha}]\ket{\Psi - \Psi _{TV}}=\sum _j A_\Psi ^{\alpha j}\langle E[\vv ^{\alpha}]\ket{c\phi_j},
\label{siple solution}
\end{equation}
where $A_\Psi ^{\alpha j}$ are linear combinations of coefficients $a_{IJK}^j$. The right-hand side can be rewritten using the normalization condition for $w^i$ as
\begin{equation}
i\sum_j2^{\Delta _{\alpha}+h_j-2} A_\Psi ^{\alpha j}\langle V^{\alpha}(i)\phi_j(0)\rangle _{\text{UHP}},
\label{simple solution 2}
\end{equation}
Substituting (\ref{bulk-boundary}) into the (\ref{coef}) one finally finds
\begin{equation}
n_{\psi}^\alpha=-\pi \sum_j2^{2h_j-1} A_\Psi ^{\alpha j}B^a_{\alpha j}.
\label{final coef}
\end{equation}

In this paper, we are mostly concerned with the Ising model. The underlying BCFT used to define particular OSFT can in the Ising sector be defined by one of three boundary states. For the $\kkett{\sigma}$ boundary condition the string field can be expanded in the basis
\begin{equation}
\Psi =\sum _{I,J,K}[a_{IJK}^0L^{I}_{-I} L^{R}_{-J}  L'^{gh}_{-K}c_1\ket0+a_{IJK}^\eps L^{I}_{-I} L^{R}_{-J} L'^{gh}_{-K}c_1\ket\eps ].
\label{field-free}
\end{equation}
while for the $\kkett{0}$ and $\kkett{\eps}$ boundary conditions the string field lives in the Verma modul of the identity
\begin{equation}
\Psi =\sum _{I,J,K}a_{IJK}^0L^{I}_{-I} L^{R}_{-J}  L'^{gh}_{-K}c_1\ket0.
\label{field-fixed}
\end{equation}
The main goal of this paper is to find such solutions in level truncation and to compute the resulting boundary states using the formula (\ref{final coef}).

\subsection{Null states}

Whereas null states are well understood and play prominent role in the conformal field theory,
they have not yet been studied in the context of string field theory.
In fact, to the contrary, in the early works on lump solutions, notably in the pioneering work \cite{MSZ}, the null states were specifically avoided by a judicious choice of parameters of the theory. It turns out however, that null states can be beneficial to string field theory numerical computations. If handled properly, they allow one to reduce the complexity of level truncation computations by reducing the number of independent components of the string field.

The basic property of null states is that they form an {\it ideal} within the string field star algebra. That means in particular, that given a null state $\eta$ and arbitrary state $\phi$ the star products $\eta * \phi$ and $\phi * \eta$ are again null. To see this, consider $\aver{\chi, \eta*\phi} = \aver{\eta,\phi*\chi}$. Since $\eta$ is null, this overlap is zero for every test state $\chi$, and hence the star product $\eta*\phi$ is null. Similarly, one can show that $\phi*\eta$ is also null. Although this seems to be a very elementary statement, it may appear surprising from certain points of view. Consider for example in the context of the Ising model $(L_{-2}-\frac{3}{4} L_{-1}^2)\ket{\eps} * \ket{\eps}$ and use the conservation laws for the $c=\half$ Virasoro operators to simplify the star product. It is not clear a priori, how this procedure should conspire to produce a sum of null states over the identity operator. In particular, the first such non-trivial null state (ignoring the descendants of the generic null state $L_{-1} \ket{0}$) should appear at level 6. Explicit computation does, however, confirm this. We find
\be
(L_{-2}-\frac{3}{4} L_{-1}^2)\ket{\eps} * \ket{\eps} =  \frac{2\times3^3}{7^2\times 11}K^{-9}\left(L_{-6} + \frac{22}{9} L_{-4} L_{-2} - \frac{31}{36} L_{-3}^2 - \frac{16}{27} L_{-2}^3 \right)\ket{0} + \cdots,
\ee
where the dots stand for higher level terms, and $K=\frac{3\sqrt{3}}{4}$ is the ubiquitous constant in string field theory raised as usual to minus the sum of the conformal weights of operators on the left and right hand sides.

Usual level truncation approach to string field theory starts with an appropriate ansatz for the string field followed by the computation of the action from which the equations of motion are derived by varying with respect to the unknowns. Had one accidently included null states in the ansatz for the string field, the action would not depend on the corresponding variable (or linear combination thereof). The equations of motion would be linearly dependent. In practise, this would be an obstacle to the efficient use of the Newton's method.

There is, fortunately, a simple systematic remedy to this problem which we use in our code. For a given basis in the matter sector, where the null states might be present, we compute level by level the Gram matrix of the inner products. If the matrix is nondegenerate, there are no null states. In the opposite case, we find the appropriate basis of non-null states (i.e. a subspace where the only null state is zero) by following the row reduction procedure from linear algebra. Permuting and subtracting rows with appropriate coefficients, we end up with the row reduced echelon (step-like) form. The columns where the leading elements (equal to 1) appear, correspond to the states that should be included in the string field. The resulting set of states depends on the chosen ordering of states, but the string field and its observables do not (barring accumulated floating point errors).

For the Ising model, there is actually more elegant option to eliminate the states. It turns out that the characters $\chi_{r,s}^{(p,p')} = q^{h_{rs}-c/24} \mathrm{ch}_{r,s}^{(p,p')}$ for the Ising model can be written in the factorized form\footnote{In general the factorizable characters are $\mathrm{ch}_{\nu,m}^{(2\nu,p')}$ and $\mathrm{ch}_{\nu,m}^{(3\nu,p')}$.} discovered by Christe in \cite{Christe} and proved by Kellendonk et al. in \cite{Kellendonk} using
\bea
\mathrm{ch}_{1,1}^{(3,4)} &=& \prod_{l \ne 0, \pm 1, \pm 8, \pm 9, \pm 10 \mod 16} (1-q^l)^{-1} \\
\mathrm{ch}_{1,2}^{(3,4)} &=& \prod_{l \ne 0, \pm 2, \pm 4, \pm 6, \pm 8 \mod 16} (1-q^l)^{-1} \\
\mathrm{ch}_{1,3}^{(3,4)} &=& \prod_{l \ne 0, \pm 2, \pm 3, \pm 5, \pm 8 \mod 16} (1-q^l)^{-1},
\eea
which suggests that it should be possible to use only the modes
\be\nonumber
L_{-2},L_{-3},L_{-4},L_{-5},L_{-11},L_{-12},L_{-13},L_{-14},L_{-18},L_{-19},L_{-20},L_{-21},\ldots
\ee
in the Verma module of the identity, and
\be\nonumber
L_{-1},L_{-4},L_{-6},L_{-7},L_{-9},L_{-10},L_{-12},L_{-15},L_{-17},L_{-20},L_{-20},L_{-22},\ldots
\ee
in the Verma module of $\eps$. In both cases the pattern repeats modulo 16. We have checked that this basis of descendants is non-degenerate numerically at least up to the level 24. Had we needed also the Verma module of $\sigma$, we would find that only $L_{odd}$ are necessary, i.e. $L_{-1}, L_{-3}, L_{-5}, L_{-7},\ldots$.

The reader might be curious to see what happens to the "full" equations of motion $Q\Psi + \Psi*\Psi=0$, if one decides {\it not} to set the null fields to zero. This is a very interesting question. While it is obvious that for every solution to the full set of equations one can find a solution of the reduced system by consistently setting all null states to zero, it is not clear whether every solution of the reduced system can be "lifted" to the full system. This is analogous to the problem studied in \cite{KMS}.  We found some little indication that this is indeed so, by performing a small little numerical experiment. For the Ising model solution discussed in section \ref{sec:identity} we observed, that if we slightly change the central charge away from $\half$ the solution still exists with almost identical coefficients, but with some definite values for the coefficients of the would be null-states. Whether every solution modulo null states can be uplifted to the full solution is an interesting issue which we postpone to the future work.

\subsection{Numerical implementation}

We perform our numerical calculations using a combination of Mathematica and C++ code. We use Mathematica for symbolic manipulations like deriving the conservation laws or computing the commutators of oscillators and C++ for time consuming tasks like computation of vertices and Newton's method. A more detailed description will appear elsewhere \cite{KudrnaSchnabl}. In this section we describe only some features specific to the Ising model.

The most efficient numerical algorithm for solving large systems of quadratic equations is the Newton's method. However, this method requires a starting point for the iteration and the final results do depend quite a lot on it. In string field theory, we do not have any a priori intuition about how good solutions should look like, so the best we can do is to start with the complete set of solutions to our system truncated to some low level, and then refine all of these starting solutions to the desired accuracy and level.

One possibility how to solve the initial lower level system of equations is using the NSolve function in Mathematica, which solves the equations using numerical algorithm based on the Gr\"obner basis. It works very well when we consider a small number of equations, but it has a bad time scaling with increasing number of equations. For example, in our test on a common desktop a system of 7 quadratic equations was completely solved in 10 seconds, 8 equations in 4 minutes, 9 in 35 minutes, 10 in 10 hours and 11 equations in 6 days. It is impossible to go much further. Not only do the memory requirements grow significantly, but also the time required grows faster than exponentially. On top of that, algorithms based on Gr\"obner basis are hard to parallelize, so this is definitely not a promising avenue.

Another option is the linear homotopy continuation method which works surprisingly well for the problem at hand. It works essentially by continuously deforming the system to a simpler one, for which all solutions are known, such as $x_i^2=1,\, \forall i$. Its advantage is much better time scaling with the number of equations and the possibility of very straightforward parallelization.

One curious feature of the system of equations of string field theory is that starting at level 2, if the system is truncated to $N$ equations in $N$ fields, it will have {\it less} than $2^N$ real or complex solutions counting possible multiplicities, which for the system coming string field theory appear to be exactly one. So the smaller number of solutions is in sharp contrast to what happens for polynomial equations in one variable, where the number of roots counting multiplicities is always given by the order of the polynomial. One can easily see it by truncating the action (\ref{action lev 2}) to $t$ and $w$ fields and solving the corresponding equations. They have only 3 solutions instead of 4. When we add the other two fields, the number of solutions becomes 15 out of possible 16. This suggests that we loose solutions when we add fields from the ghost sector, which is also consistent with our calculations up to level~4.


When using the linear homotopy method for larger set of equations, one cannot be absolutely sure without further analysis whether one finds all the solutions. The $2^N$ solutions of the deformed system can merge under the homotopic continuation and the question is whether it is because of numerical accuracy or because it is a true feature of the system. One can exclude the former possibility beyond reasonable doubt by playing with the parameters such as the step-size in the homotopy and the numerical precision.


Another issue we would like to mention is an unpleasant loss of numerical precision we have encountered during the search for solutions using Newton's method. We stop the iterations of Newton's method when $\frac{\|\psi_{(i)} -\psi_{(i-1)}\|}{\|\psi_{(i)}\|}<p$, where $\|\ \|$ represents quadratic norm, $\psi_{(i)}$ is the solution in the $i$-th iteration and $p$ is the target precision, which we mostly take to be $10^{-12}$. In other string field theory models we found that we can reach the precision of the number format we use, but this is not true for Ising model. Concretely on the  $\mathds{1}$-brane at level 16 we find that we cannot reach the $10^{-12}$ precision using double format of numbers in C++. Therefore we changed the number format to long double, but at level 24 we encountered this problem again, so we were forced to reduce the target precision by one order.
This phenomenon is general for all solutions in the same ansatz including tachyon vacuum. The exact precision we can get slightly depends on the solution and the representation of the irreducible basis in the Ising sector. Curiously we did not encounter this problem on the $\sigma$-brane, but it can be only because we did not go to high enough level.
Since the problems disappear when we slightly change the central charge to a nonphysical value, where there are no null states in the spectrum, we have to conclude that this instability is caused by the numerical removal of the null states.


\section{Solutions on the $\sigma$-brane}
\label{sec:sigma}
\setcounter{equation}{0}

In this section, we will start with the Ising model on the the upper half plane with free boundary conditions, given by the boundary state $\kkett{\sigma}$, and by studying solutions of OSFT formulated for this background we will show how to reach other possible boundary conditions.
In this setup the string field can be written in the form (\ref{field-free}) truncated to some level, which is defined as the eigenvalue of the $L_0^{\mathrm{tot}}+1$ operator, where $L_0^{\mathrm{tot}}=L_0^{\mathrm{matter}}+L_0^{\mathrm{ghost}}$ is the total Virasoro generator.
\begin{figure}
        \centering
        \begin{subfigure}[t]{0.47\textwidth}
                \includegraphics[width=\textwidth]{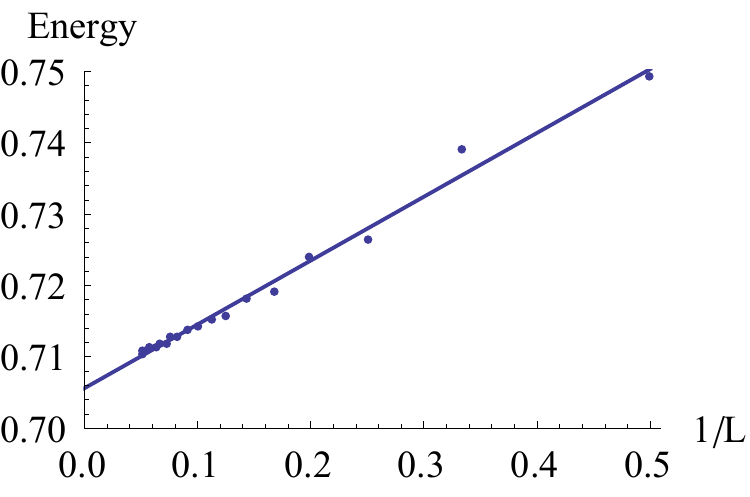}
        \end{subfigure}\qquad
        \begin{subfigure}[t]{0.47\textwidth}
                \includegraphics[width=\textwidth]{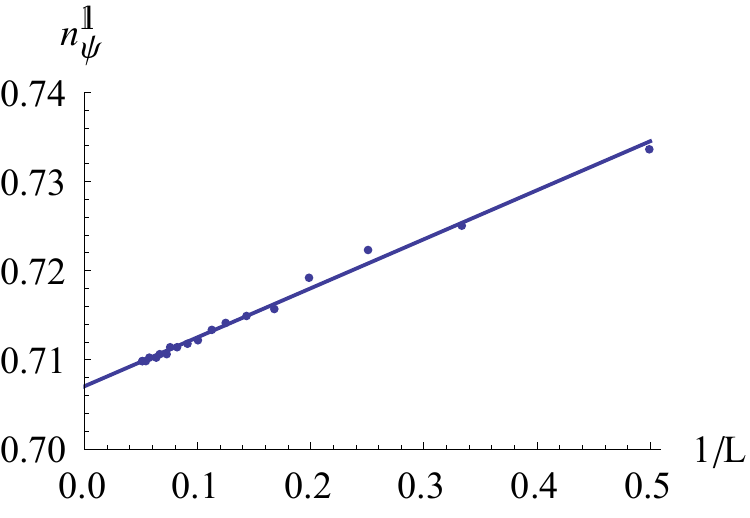}
        \end{subfigure}

        \begin{subfigure}[t]{0.47\textwidth}
                \includegraphics[width=\textwidth]{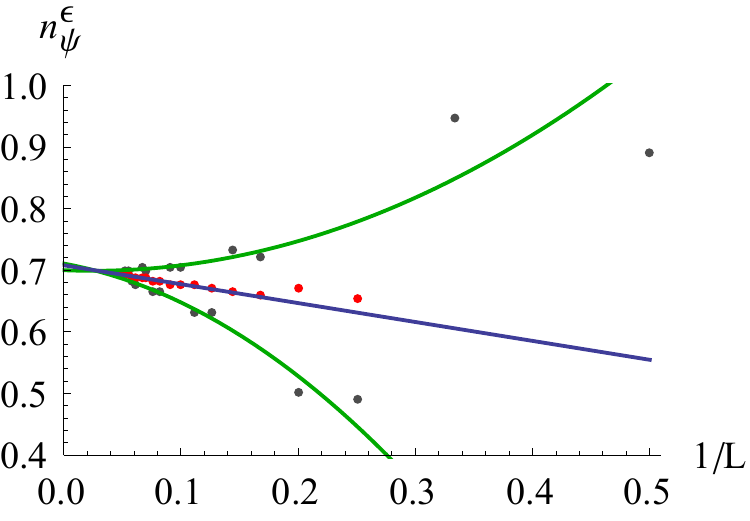}
        \end{subfigure}\qquad
        \begin{subfigure}[t]{0.47\textwidth}
                \includegraphics[width=\textwidth]{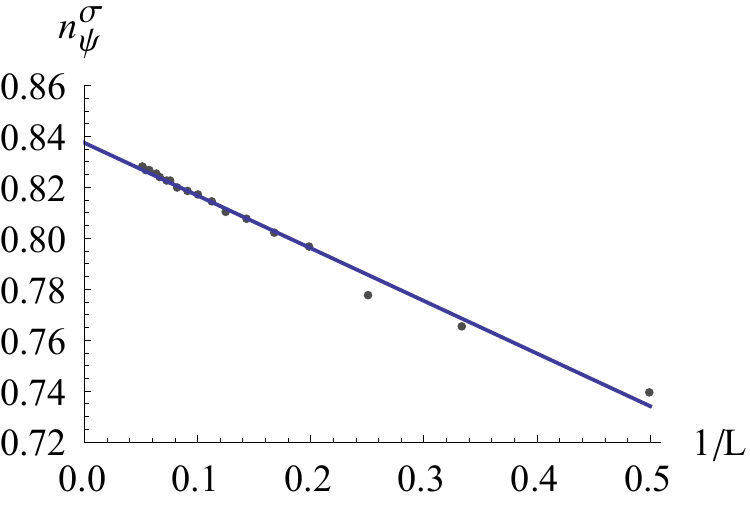}
        \end{subfigure}
        \caption{Gauge invariants for the solution $\Psi$ corresponding to the $\mathds{1}$-brane found in the $\sigma$-brane background. For all invariants a linear extrapolation to the infinite level is visualized. Due to huge oscillations for the invariant $n_\Psi^\eps$, we also added its Pad\'e-Borel approximation to smoothen the data (red). Moreover, we performed two quadratic fits (green) using the two branches of the oscillating data. We can see quite good agreement for all the three extrapolations.}\label{sigma_brane}
\end{figure}

Let us illustrate our method by truncating the string field to the lowest non-trivial level $\half$ \begin{equation}
\ket\Psi =t c_1\ket0 + a c_1\ket\eps.
\label{level 0.5 field}
\end{equation}
and let us compute the OSFT action following \cite{MSZ}.
The kinetic term is readily found
\begin{eqnarray}
\frac{1}{2} \VV\Psi{ Q\Psi}&=&\frac{1}{2} \VV{tc_1 + ac_1\eps}{c_0L_0[tc_1 + ac_1\eps]}=\nonumber\\
&=&-\frac{1}{2}t^2\bra0 c_{-1}c_0c_1\ket0 -\frac{1}{4}a^2\bra\eps c_{-1}c_0c_1\ket\eps =-\frac{1}{2}t^2-\frac{1}{4}a^2,
\label{kinetic term}
\end{eqnarray}
where we have used the knowledge of the boundary 2-point functions, normalization of the ghost fields $\bra0 c_{-1}c_0c_1\ket 0=1$, and the identity $\langle \Psi ,Q\Psi \rangle = \langle \Psi ,c_0L_0\Psi \rangle$ valid for all fields in the Siegel gauge. 
For the interaction term, we find
\begin{eqnarray}
\frac{1}{3} \VVV\Psi\Psi\Psi =\frac{1}{3}t^3\langle c_1,c_1,c_1 \rangle + \frac{1}{3}3a^2t\langle c_1 \eps ,c_1 , c_1 \eps \rangle=\frac{1}{3}K^3t^3 +K^2a^2t,
\label{interaction term}
\end{eqnarray}
where the factor 3 in the second step corresponds to the three ways of inserting the $\eps$ operators. The often appearing ghost contribution is denoted as usual by $\langle c_1,c_1,c_1\rangle = K^3=\big (\frac{3\sqrt{3}}{4}\big )^3$. In the computation we have used the three-point function (\ref{three-point}) with the corresponding structure coefficients for the identity and the $\eps$ operators. In total, the OSFT action truncated to the level $\half$ for the string field (\ref{level 0.5 field}) is
\begin{eqnarray}
\mathcal{V} (t,a)  = -\frac{1}{2}t^2-\frac{1}{4}a^2+\frac{1}{3}K^3t^3 +K^2a^2t.
\label{truncated action}
\end{eqnarray}
Minimizing this action, we find two solutions corresponding to the two other elementary boundary states $t$=0.14815, $a$=$\pm$0.24348 for which the normalized energy difference between the tachyon vacuum and the solution is $2\pi ^2 [\mathcal{V}(\Psi)-\mathcal{V}(\Psi _{TV})]$= 0.83029 which differs only by 17.4 $\%$ from the expected value $\frac{1}{\sqrt{2}}$ (i.e. overlap of the corresponding boundary state with $|0\rangle$, see table \ref{tab:Ising-branes}).

Now, using formula (\ref{final coef}) let us compute Ellwood invariants for bulk operator $i = \mathds{1}$ to illustrate a typical computation. For the string field (\ref{level 0.5 field}) truncated to the level $\half$, we get
\begin{equation}
 n^\mathds{1}_\Psi = -\frac{\pi}{2}\Big[t\langle \mathds{1}\rangle _{UHP}+a\langle \eps (0)\rangle _{UHP}\Big]+1=-\frac{\pi}{2}t+1,
\end{equation}
where the additive constant corresponds to $\Psi _{TV}$ term and ensures correct normalization. For the two solutions we obtain numerical value $ n^\mathds{1}_\Psi =$ 0.76729 which differs only by 8.5 $\%$ from the expected value of the coefficient in front of the corresponding Ishibashi state $\kett {0}$ in (\ref{boundary states}). Similar procedure can repeated for the other two Ellwood invariants and we get $ n^\eps _\Psi = -0.76729$  and  $n^\sigma_\Psi = 0.64320$. We can see a big disagreement in $ n^\eps _\Psi$ coefficient but things will go better as we move to level 2.

Level 2 is the next nontrivial level, where descendants of the identity appear. The string field truncated to this level takes the form
\begin{equation}
\ket\psi =tc_1\ket0 + ac_1\ket\eps+uL'^{gh}_{-2}c_{1}\ket0+vc_1L^I_{-2}\ket0+wc_1L^R_{-2}\ket0,
\label{level 2 field}
\end{equation}
where some fields which can be consistently set to zero by symmetries have been omitted. To find the action, requires just a little bit more work, still manageable by hand. The conservation laws of \cite{RZ} are useful to deal with the Virasoro descendants. The resulting action \cite{Rapcak} has in addition to the perturbative and tachyon vacua two interesting critical points $t$=0.21084, $a$=$\pm$0.27990, $u$=-0.02703, $v$=-0.09947, $w$=0.0301 with energy 0.75421 that approaches the expected value and differs only by 6.7~$\%$. To compute the Ellwood invariants, it is very convenient to use the conservation laws \cite{KKT,KMS} such as 
\begin{equation}
\bra{ E[\mathds{1}]}L_{-2}^I=\bra{ E[\mathds{1}]}L_{2}^I+\frac{c_I}{2}\bra{ E[\mathds{1}]}.
\end{equation}
and similarly for the other bulk fields. If we insert the coefficients for the solutions we find the value $ n^\mathds{1}_\Psi = 0.73370$ differing only by 3.8 $\%$ from expected value. Moreover, we can find  $ n^\eps_\Psi = 0.89339$ that is much closer to the expected value than in the level $\half$ case. General feature of the computation of the boundary states using level truncation method is decreasing convergence for coefficients corresponding to the Ishibashi states associated with higher dimensional primaries.

The outlined procedure can be easily automatized and we can obtain and analyze higher level solutions. Using our C++ code we have obtained solutions for the two boundary states discussed above up to the level 20. The results for the energy and the Ellwood invariants providing the boundary state coefficients at increasing levels are given in table \ref{sol-sigma brane}. To get more accurate results, we can extrapolate the results to the infinite level. The simplest approach is to use the linear fits in $1/L$ (see figure \ref{sigma_brane}) which work already quite well. We have fitted every integer level in the case of $n_\Psi^{\mathds{1}}$ and $n_\Psi^{\sigma}$ coefficients. Coefficient $n_\psi ^{\mathds{1}}$ agrees with the expected value within 0.01 $\%$. The results for the $n_\Psi^{\eps}$  coefficient have slower convergence, there are visible oscillations (with a decreasing amplitude) with a period of four levels. In this case there are two reasonable options to proceed. One possibility is to smooth out oscillations by Pad\'{e} or Pad\'{e}-Borel resummation \cite{ErlerSch}. The latter one predicts value 0.69925 which is within 1\% from the correct answer. Another possibility is to use quadratic extrapolations for the even and odd data points separately. These two extrapolations lead to values 0.71078 and 0.70000. Their mean value is then 0.70539 and so the difference from the expected value is only 0.2 $\%$.

\begin{table}[ht]
\centering
\begin{tabular}{| r| l l l l |}
\hline
& & & & \\[-2ex]
$\mbox{Level}$    &  Energy    & $ n_\Psi ^{\mathds{1}}$  &  $n_\Psi ^{\eps }$    & \hspace{3mm} $n_\Psi ^{\sigma }$                   \\   \hline
2   & 	0.74917   & 	0.73370   & 	0.89339   & 	$\pm$ 0.73942	 \\
4   & 	0.72656   & 	0.72213   & 	0.48762   & $\pm$	0.77824	 \\
6   & 	0.71933   & 	0.71585   & 	0.72112   & $\pm$	0.80182	 \\
8   & 	0.71596   & 	0.71401   & 	0.62984   & $\pm$	0.81011	 \\
10   & 	0.71404   & 	0.71216   & 	0.70480   & $\pm$	0.81679	 \\
12   & 	0.71280   & 	0.71154   & 	0.66492   & $\pm$	0.82018	 \\
14   & 	0.71193   & 	0.71065   & 	0.70130   & $\pm$	0.82331	 \\
16   & 	0.71129   & 	0.71035   & 	0.67919   & $\pm$	0.82517	 \\
18   & 	0.71080   & 	0.70983   & 	0.70060   & $\pm$	0.82699	 \\
20   & 	0.71043   & 	0.70978   & 	0.69080   & $\pm$	0.82815	 \\   \hline
$\infty$  & 0.70560  & 0.70703  & 0.70539 & $\pm$ 0.83744 \\   \hline
$\infty _{PB}$  & -  & -  & 0.69925 & - \\   \hline
$\mbox{Expected}$   & 0.70711   & 	0.70711   & 	0.70711   & 	$\pm$ 0.84090	 \\   \hline
  \end{tabular}
\caption{Boundary coefficients for the two solutions on the $\sigma$-brane computed at even levels up to $L=20$ together. We show also the extrapolated values (as described in the main text) and exact values coming from the $\kkett{\mathds{1}}$ and $\kkett{\eps}$ boundary states.}
\label{sol-sigma brane}
\end{table}

\section{Solutions on $\mathds{1}$-brane and $\eps$-brane}
\label{sec:identity}
\setcounter{equation}{0}

In this section we look for solutions on the $\mathds{1}$-brane or $\eps$-brane, however since these two BCFTs differ only by one sign in the bulk correlation functions we will choose the $\mathds{1}$-brane description. The $\mathds{1}$-brane has the lowest possible energy $\frac{1}{\sqrt{2}}$, therefore we have to look for solutions with zero or positive energy.

In the spectrum we have only the Verma module of identity, which makes the system simpler than on the $\sigma$-brane, but it is more difficult to find nontrivial solutions. The only relevant operator available is $c_1\ket{0}$, which means we have to go at least to level 2 to find a solution different from the tachyon vacuum, as opposed to the $\sigma$-brane, where the solutions are created by exciting $c_1\ket{\eps}$. The string field truncated to level 2 has the form
\begin{equation}
\ket{\Psi}=tc_1\ket{0}+uL_{-2}^Rc_1\ket{0}+vL_{-2}^Ic_1\ket{0}+wL'^{gh}_{-2}c_1\ket{0}
\end{equation}
and the action is
\begin{eqnarray}\label{action lev 2}
S=&&-\frac{1}{2}t^2+\frac{27\sqrt{3}}{64}t^3-\frac{765\sqrt{3}}{256}t^2u+\frac{51}{8}u^2+\frac{39083}{1024\sqrt{3}}tu^2-\frac{16616021}{331776\sqrt{3}}u^3 \\
\nonumber&&-\frac{15\sqrt{3}}{256}t^2v+\frac{425}{512\sqrt{3}}tuv-\frac{195415}{110592\sqrt{3}}u^2v+\frac{1}{8}v^2+\frac{1049}{3072\sqrt{3}}tv^2-\frac{89165}{110592\sqrt{3}}uv^2\\
\nonumber&&+\frac{1427}{12288\sqrt{3}}v^3+\frac{33\sqrt{3}}{64}t^2w-\frac{935}{128\sqrt{3}}tuw+\frac{429913}{27648\sqrt{3}}u^2w-\frac{55}{384\sqrt{3}}tvw+\frac{4675}{13824\sqrt{3}}uvw \\
\nonumber&&+\frac{11539}{82944\sqrt{3}}v^2w-\frac{1}{2}w^2+\frac{19}{64\sqrt{3}}tw^2-\frac{1615}{2304\sqrt{3}}uw^2-\frac{95}{6912\sqrt{3}}vw^2+\frac{1}{64\sqrt{3}}w^3.
\end{eqnarray}

When we solve the corresponding equations of motion there is no reasonable nontrivial real solution, but we find the following complex solution: $t=0.0338302-0.312394i$, $u=1.19036+0.526409i$, $v=0.0245283-0.042421i$, $w=0.0253244-0.123227i$ and its complex conjugate. Notice that the non-diagonal primary $L_{-2}^R-51L_{-2}^I$ is excited.

Although the solution seems to be pretty wild at the first sight, we find that it is stable under level truncation. The imaginary part of the solution is getting smaller as we increase the level and surprisingly it disappears completely at level 14. We were able to evaluate the solution up to level 24, the data are shown is shown in table \ref{tab:sigma-on-1}. The last column shows the ratio between the norm of the imaginary and real parts of the solution\footnote{This ratio is computed as $\sum_i \rm{Im}[t_i]/\sum_i \rm{Re}[t_i]$, where $t_i$ are the components of the string field. It is not an invariant quantity, but it gives a good idea how big the imaginary part of the solution is.}.
\begin{table}\nonumber
\centering
\footnotesize{
\begin{tabular}{|l|lllll|}\hline
Level         & Energy        & $n_\Psi^\mathds{1}$    & $n_\Psi^\eps$    & $n_\Psi^\sigma$       & Im/Re   \\\hline
2         &  $1.59267+0.72688i$ & $1.06048  -0.18455i$  & $-9.73471-5.23904i $  & $-0.34358-0.97082i$ & 0.78840 \\
4         &  $1.41414+0.20152i$ & $0.96290  -0.14267i $ & $-0.66854+1.99191i$  & $-0.36976-0.56423i$ & 0.43838 \\
6         &  $1.28579+0.07668i$ & $0.92262   -0.11378i $ & $-3.86207-0.37376i $  & $-0.38933-0.39436i$ & 0.30746 \\
8         &  $1.21160+0.03054i$ & $0.90480   -0.08685i $ & $-0.57514+0.82266i $  & $-0.37217-0.28194i$ & 0.22100 \\
10        & $1.16345+0.01007i$ & $0.89256   -0.06174i $ & $-2.48552+0.00261i $  & $-0.37629-0.19232i$ & 0.15222 \\
12        & $1.12943+0.00123i$ & $0.88510   -0.03109i $ & $-0.56951+0.24561i$   & $-0.36891-0.09399i$ & 0.07487\\
14        & $1.10568          $        & 0.91469                   & $-1.93951          $ & $-0.26607      $     & 0        \\
16        & $1.09045        $          & 0.93044                   & $-0.95087        $  & $-0.20633      $     & 0        \\
18        & $1.07936      $            & 0.93918                   & $-1.69824      $     & $-0.17497     $      & 0        \\
20        & $1.07084    $              & 0.94538                   & $-1.04849    $       & $-0.15003     $       & 0        \\
22        & $1.06405  $                & 0.94994                   & $-1.55407  $         & $-0.13398     $       & 0        \\
24        & $1.05850$                  & 0.95367                   & $-1.08669$           & $-0.11918    $       & 0        \\ \hline
\rm{Expected} & 1                  & 1                    & $-1$                 & \hspace{0.16cm} 0                   &         \\ \hline
\end{tabular}
}
\caption{Convergence of coefficients in $\kkett{\sigma}$ found as a solution on the $\mathds{1}$-brane.}
\label{tab:sigma-on-1}
\end{table}

\begin{figure}
        \centering
        \begin{subfigure}[t]{0.47\textwidth}
                \includegraphics[width=\textwidth]{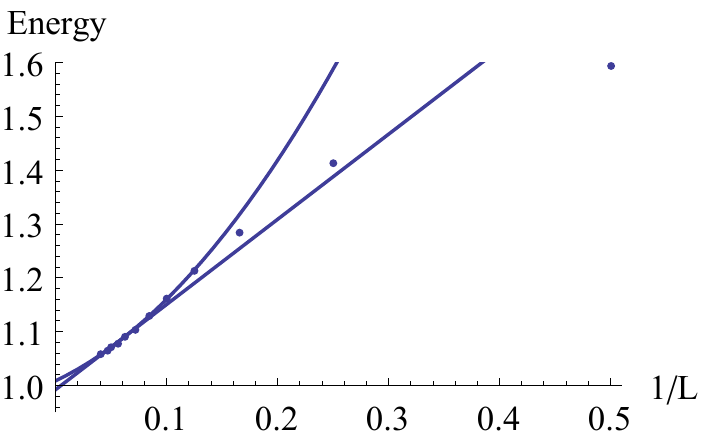}
                \caption{The energy: Unlike the Ellwood invariants the energy is relatively smooth with respect to the level, which may be caused by its quadratic dependence on the string field.}
                \label{fig:1_E}
        \end{subfigure}\qquad
        \begin{subfigure}[t]{0.47\textwidth}
                \includegraphics[width=\textwidth]{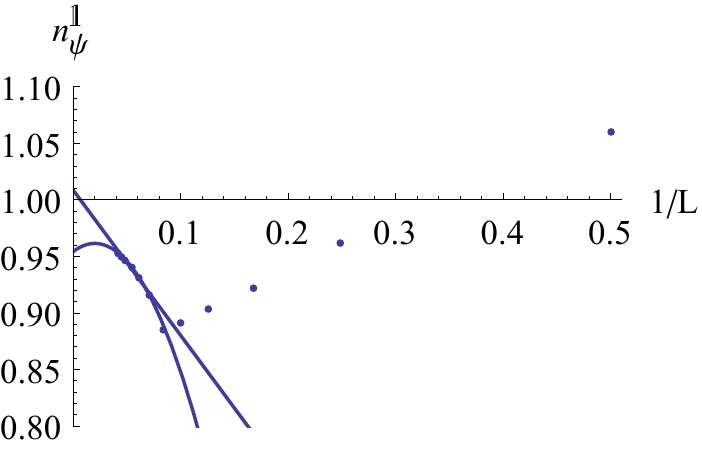}
                \caption{Invariant $n_\Psi^\mathds{1}$: Notice that at level $L$=14 the behavior changes dramatically. The linear fit works better in the case of this invariant.}
                \label{fig:1_1}
        \end{subfigure}

        \begin{subfigure}[t]{0.47\textwidth}
                \includegraphics[width=\textwidth]{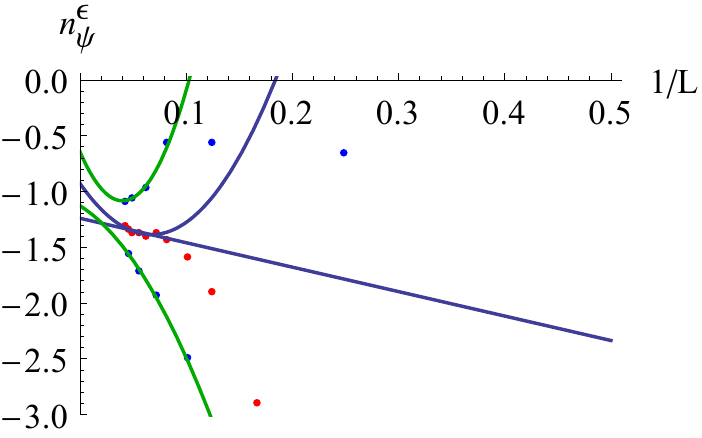}
                \caption{Invariant $n_\Psi^\eps$ (blue):  Due to the oscillations, we also added its Pad\'e-Borel approximation to smoothen the data (red). The two fits are performed using these modified data. Moreover, we performed two quadratic fits using the two branches of the oscillating data. The overall agreement is good at best at the qualitative level.}
                \label{fig:1_epsilon}
        \end{subfigure}\qquad
        \begin{subfigure}[t]{0.47\textwidth}
                \includegraphics[width=\textwidth]{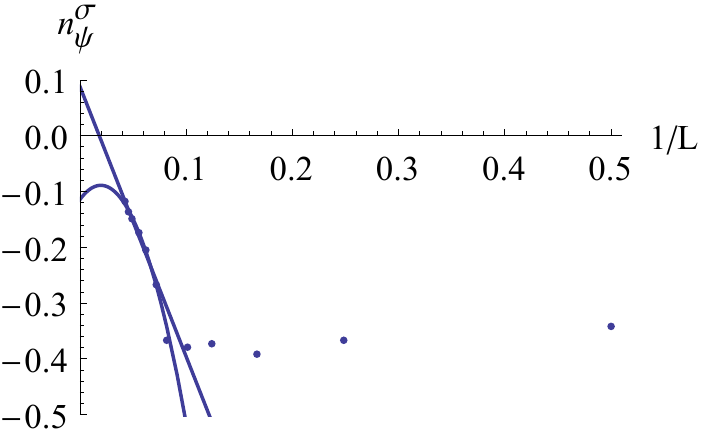}
                \caption{Invariant $n_\Psi^\sigma$: Up to the level $L$=14 the invariant is nearly constant, then its starts to grow towards zero. Unfortunately, neither one of the the fits captures the data very well.}
                \label{fig:1_sigma}
        \end{subfigure}
        \caption{Real parts of gauge invariants for the solution $\Psi$ corresponding to the $\sigma$-brane found in the $\mathds{1}$-brane background. For all invariants a linear and quadratic extrapolation to the infinite level is visualized.}\label{identity_brane}
\end{figure}

The energy of the solution is close to one, so the most likely interpretation is a $\sigma$-brane. The components of boundary state do roughly agree with the expected values (\ref{boundary states}), but the agreement is much worse than in the previous section. The dependence of the real part of the energy and of the invariants on $1/L$ is plotted in figures \ref{fig:1_E} - \ref{fig:1_sigma}.

From the figures we can see that the behavior of the invariants change drastically at level 14. Up to this level the real parts of the invariants are going in the wrong direction, but from there on they start to converge to the correct values. At level 24 the dependence on the level is still quite significant, so to get a reliable estimate we should make an extrapolation to the infinite level. The sharp change in the behavior at level 14 suggests that for reliable fits (especially of low orders) we should exclude the results from lower levels. Then only relatively few data points remain, and to maintain certain degree of predictability we study only linear and quadratic fits in $1/L$. The difference in the asymptotic value between these fits can be taken as some sort of an estimate for the systematic error. For the $n_\psi^\eps$ invariant, the values oscillate significantly so to obtain something at least remotely meaningful, we smoothened the data. We tried both the Pad\'e-Borel approximation and separately treating the level 0 and 2 mod 4. The results of the Pad\'e-Borel approximation are in the following table:
\begin{table}\nonumber
\centering
\begin{tabular}{|l|l|}\hline
Level  & $n_{\Psi\ PB}^\eps$ \\\hline
4  & $-4.36484-0.74899i$ \\
6  & $-2.90862-0.10512i$ \\
8  & $-1.88644+0.33130i$ \\
10 & $-1.57741+0.24252i$ \\
12 & $-1.43600+0.26176i $   \\
14 & $-1.38446 $\\
16 & $-1.38544 $\\
18 & $-1.37157 $\\
20 & $-1.35608 $\\
22 & $-1.33856 $\\
24 & $-1.31994$ \\\hline
\end{tabular}
\caption{Values of $n^\eps$ obtained by the Pad\'{e}-Borel approximation.}
\end{table}

The fitter curves are shown at figures \ref{fig:1_E} - \ref{fig:1_sigma} and the extrapolated values of energy and components of the boundary state are
\begin{equation}\nonumber
\begin{array}{|l|l|l|l|l|}\hline
\rm{fit}       & \rm{Energy}^{(\infty)}  & n_\Psi^{\mathds{1}(\infty)} & n_\Psi^{\eps(\infty)} & n_\Psi^{\sigma(\infty)} \\\hline
\rm{linear}    & 0.9920                  & 1.0090                      & -1.24                   & 0.0882 \\\hline
\rm{quadratic} & 1.0085                  & 0.9540                      & -0.9269                   & -0.1147 \\\hline
\end{array}
\end{equation}
Unfortunately the linear and quadratic extrapolation give significantly different results and we have also tested that the extrapolated values change a lot when we add or remove a single data point, so we have to conclude that the fits have large errors and more data will be needed to make the extrapolations really trustworthy.

The energy is approximately within 1\% from the expected value, which is a very good agreement, but the other invariants are mostly quite far off the correct values. However the expected boundary state always lies between the two extrapolations and the invariants move in the right direction with increasing level, so we believe that the solution is really a $\sigma$-brane.

In order to find the $\eps$-brane or multi-brane solutions we have also found starting points at level 4. There is over 8000 starting solution, but when we improved them to level 16 and removed all nonstable and duplicate or complex conjugated solutions, only about 250 solutions remained. Most of them have action which is too high, negative or inconsistent with the $n_\psi^\mathds{1}$ invariant, but we present few of the interesting ones in appendix \ref{ap: solutions}.

In the end we emphasize that all solutions constructed on the $\mathds{1}$-brane are also consistent solutions on the  $\sigma$-brane, because they contain only descendants of the identity operator. When these solutions are interpreted on the $\sigma$-brane the energy and the invariants change because of the overall change in the normalization of correlation functions. The energy and $n_\psi^\mathds{1}$ invariant gets multiplied by $\sqrt{2}$, the $n_\psi^{\eps}$ invariant by $-\sqrt{2}$ and $n_\psi^\sigma$ by zero. These values are given in table \ref{tab:reint-sigma-on-1} and the interpretation of this solution is clearly the boundary state $\kkett{\mathds{1}}+\kkett{\eps}$. This nicely illustrates the general formula (\ref{boundary states relation}) discussed in the introduction. 

\begin{table}\nonumber
\centering
\begin{tabular}{|l|llll|}\hline
Level         & Energy        & $n_\Psi^\mathds{1}$  & $n_\Psi^\eps$    & $n_\Psi^\sigma$\\\hline
2         & $2.25238+1.02796i$   & $1.49975-0.26099i $ & $13.7670+7.40912i $   & 0 \\
4         & $1.99989+0.28499i$  & $1.36175-0.20177i $ & $0.94546-2.81698i $ & 0 \\
6         & $1.81839+0.10844i$  & $1.30478-0.16091i $ & $5.46180+0.52857i$   & 0 \\
8         & $1.71346+0.04319i$ & $1.27958-0.12283i $ & $0.81337-1.16342i$  & 0 \\
10        & $1.64537+0.01424i$ & $1.26228-0.08731i$ & $3.51506-0.00369i$ & 0 \\
12        & $1.59725+0.00173i$ & $1.25172-0.04397i$ & $0.80541-0.34734i$ & 0 \\
14        & 1.56367            & 1.29357            & 2.74288            & 0 \\
16        & 1.54212            & 1.31585            & 1.34474            & 0 \\
18        & 1.52644            & 1.32820             & 2.40168            & 0 \\
20        & 1.51439            & 1.33698            & 1.48279            & 0 \\
22        & 1.50480             & 1.34342            & 2.19779            & 0 \\
24        & 1.49695            & 1.34869            & 1.53682            & 0 \\ \hline
\rm{Expected} & 1.41421            & 1.41421            & 1.41421            & 0 \\ \hline
\end{tabular}
\caption{Reinterpretation of the solution found on the $\mathds{1}$-brane as a solution in the $\sigma$-background and corresponding rescaled boundary state coefficients.}
\label{tab:reint-sigma-on-1}
\end{table}

\section{Double Ising model conformal boundary conditions}
\label{sec:double}
\setcounter{equation}{0}

\subsection{Quick review of D-branes in (Ising)$^2$ model}
Similarly as in the case of the Ising model we can construct boundary states in tensor product of two Ising model CFT's. It is interesting to consider boundary states in such tensored models since they they are in correspondence with defects in the simple model. Moreover, this model is dual to the free boson living on the orbifold $S^1/Z_2$ with radius $\sqrt{2}$, in units where $\alpha'=1$.

\begin{table}[H]
\centering
\begin{tabular}{|l|c|l|l|}\hline
& & &\\[-1.5ex]
$h=\bar h$  & $\mathrm{Multiplicity}$ & $(\mathrm{Ising})^2 \;\; \mathrm{Examples}$ & $\mathrm{Orbifold \;\; Examples}$\\[0.5ex]\hline
& & & \\[-2ex]
$\multirow{2}{*}{$n^2=0,1,4,\ldots$} $  & $\multirow{2}{*}{1}$  & $\mathds{1}\otimes \mathds{1}$  & $\mathds{1}$ \\
& & $\eps\otimes\eps$  & $-2\partial X \bar\partial X$ \\[0.5ex]
$\multirow{2}{*}{$\frac{(n+1)^2}{2}=\frac{1}{2},2,\frac{9}{2},\ldots$}$ & $\multirow{2}{*}{2}$ & $\frac{1}{2}(\mathds{1}\otimes\eps+\eps\otimes\mathds{1})$ & $\cos(\sqrt{2} X)$\\
&  & $\frac{1}{2}(\mathds{1}\otimes\eps-\eps\otimes\mathds{1})$ & $\pm\cos(\sqrt{2} \tilde X)$\\[1ex]
$\frac{(2n+1)^2}{8}=\frac{1}{8},\frac{9}{8},\frac{25}{8},\ldots $ & 1 & $\sigma\otimes\sigma$ & $\pm \sqrt{2}\cos(\frac{X}{\sqrt{2}})$ \\[1ex]
$\multirow{2}{*}{$\frac{(2n+1)^2}{16}=\frac{1}{16},\frac{9}{16},\frac{25}{16},\ldots$}$ & $\multirow{2}{*}{2}$ & $\mathds{1}\otimes\sigma , \sigma\otimes \mathds{1}$ & $\mathrm{twist \; fields}$\\
& & $\eps \otimes\sigma, \sigma\otimes\eps $&$ \mathrm{excited\; twist \; fields}$ \\[0.5ex]\hline
\end{tabular}
\caption{Spinless primary fields in the double Ising model with low-level examples and their free boson orbifold duals.}
\label{primaries}
\end{table}

It is clear that tensor products of boundary states of the Ising model remain to be boundary states in the double Ising model. There are nine such product boundary states but these are not all the boundary states of the double Ising model. The situation becomes more complicated since infinite set of new  spinless primary fields such as $(\bar{L}_{-2}^{(1)}-\bar{L}_{-2}^{(2)})(L_{-2}^{(1)}-L_{-2}^{(2)})\mathds{1}$ and corresponding Ishibashi states appear. These primary fields belong to one of the four infinite towers (two of them with multiplicity 2) found in \cite{yang}. For readers convenience we list them in table \ref{primaries} and provide also the duals of the lowest lying states under the orbifold correspondence. For more details see the appendix \ref{ap: double_ising}.

The boundary states have been found by Affleck and Oshikawa \cite{Oshikawa} and we review them in the table \ref{branes}. Using correspondence of primary fields from table \ref{primaries} and the explicit knowledge of boundary states of the Ising model, we are able to interpret boundary states in the orbifold picture. We can see that eight of the nine tensored boundary states form fractional branes of the orbifold boson whereas the last boundary state corresponds to the centered bulk D0-brane. This last boundary state is in fact a member of a continuous family of boundary states, since the generic D0-brane is free to move along the orbifold. Finally, there is one last continuous family of boundary states which corresponds to the generic D1-brane with an arbitrary Wilson line turned on its worldvolume. Detailed discussion of the boundary states and corresponding boundary spectra can be found also in appendix \ref{ap: double_ising}.

\begin{table}[H]
\centering
  \begin{tabular}{|l|c|c|c|c|c|c|}\hline
& & & & & &\\[-1.5ex]
 $(\mbox{Ising})^2$ & $\mathrm{Interpretation}$ &  $\mathrm{Energy}$ &  D0/D1 &$\mathrm{Position}$&  T-dual position         & Twist charge\\[-0.2ex]
 D-brane & $ $ &  $ \aver{\mathds{1}}$ &  $\frac{\aver{\partial X\bar\partial X}}{\aver{\mathds{1}}} $ &$\langle X\rangle $& \centering $\langle \tilde{X}\rangle $         &\\[0.5ex]  \hline
 & & & & &  &\\[-2ex]
 $\mathds{1} \otimes \eps $& $\mathrm{fractional\;D0 }$ & $\frac{1}{2}$ & +1 & $\pi R$ &   -      & $+1$ \\[0.5ex]
 $\eps  \otimes \mathds{1}$& $\mathrm{fractional\;D0 }$ & $\frac{1}{2}$ & +1 & $\pi R$ &    -   &$-1$ \\[0.5ex]
 $\mathds{1} \otimes \mathds{1}$ & $\mathrm{fractional\;D0 }$ & $\frac{1}{2}$ & +1  & 0 &  -     &$+1$\\[0.5ex]
 $\eps \otimes \eps $ & $\mathrm{fractional\;D0 }$ & $\frac{1}{2}$ & +1  & 0 &    -   &$-1$\\[0.5ex]
 \hline
  & & & & &       &\\[-1.5ex]
$\mathds{1} \otimes \sigma $ & $\mathrm{fractional\;D1 }$ & $\frac{1}{\sqrt{2}}$ & $-1$ & - &    $\frac{\pi}{R}$    &+1 \\[0.5ex]
 $\sigma  \otimes \mathds{1}$& $\mathrm{fractional\;D1 }$ & $ \frac{1}{\sqrt{2}} $& $-1$ & -  &     0  & +1\\[0.5ex]
 $\eps \otimes \sigma $ &$ \mathrm{fractional\;D1 }$ & $\frac{1}{\sqrt{2}}$ & $-1$ &  - &    $\frac{\pi}{R}$   &$-1$\\[0.5ex]
 $\sigma \otimes \eps$ & $\mathrm{fractional\;D1 }$ & $\frac{1}{\sqrt{2}}$ & $-1$ & - &    0   &$-1$\\[0.5ex]
 \hline
 & & & & &       &\\[-1.5ex]
 $\sigma \otimes \sigma$ & $\mathrm{centered\; bulk \;D0 }$ & 1 & +1 & $\frac{\pi R}{2}$  &   $\frac{\pi }{2R}$    &0\\[0.5ex]
 \hline
  & & & & &       &\\[-1.5ex]
$\kkett{D_O(\phi)}$ & $\mathrm{generic\; bulk \;D0 }$ & 1 & +1 & $\phi R $ &   -    &0\\[0.5ex]
 \hline
 & & & & &       &\\[-1.5ex]
$\kkett{N_O(\tilde{\phi})}$ & $\mathrm{generic\; bulk \;D1 }$ & $\sqrt{2}$ & $-1$ & - &   $\frac{\tilde{\phi}}{R}$    & 0\\[0.5ex]
 \hline
 \end{tabular}
\caption{D-branes in the double Ising model and their free boson duals with few parameter describing them.}
\label{branes}
\end{table}

The solution on the $\mathds{1}$-brane corresponding to $\sigma$-brane from section \ref{sec:identity} can be also extended to double Ising model by letting the second Ising sector (for example also with $\mathds{1}$-brane boundary condition) be part of the universal sector in the setup of section \ref{sec:identity}. The resulting solution can be interpreted as fractional D0-brane being excited into the fractional D1-brane. The same solution on $\sigma\ot \mathds{1}$-brane can be interpreted as a fractional D1-brane changing into two fractional D0-branes located at the two invariant orbifold points.

The components of the boundary states should satisfy (\ref{boundary states relation}). However we find that this is true only in a specific basis of operators. On the Ising model side of the duality the equation is satisfied in a factorized basis of $\mathds{1}\ot \eps,\ \eps\ot \mathds{1}$ (should be verified for higher levels) and on the orbifold side only when we consider a strange basis $\cos \sqrt{2}X+\cos \sqrt{2}\tilde X,\ \cos \sqrt{2}X-\cos \sqrt{2}\tilde X$. In the usual basis the equation (\ref{boundary states relation}) is valid only when we mix winding and momentum components of the boundary states. This is probably related to the fact that the boundary spectrum on the fractional D0-brane contains only winding modes and on the fractional D1-brane only momentum modes. Therefore when reinterpreting the solution we have to move it from one Verma module to another, although they have the same weight.


\subsection{Numerical results}

As stated above, tensor products of the original Ising model boundary states remain to be boundary states in the doubled Ising model. In the following, we will restrict ourselves to these products. Starting with a background $\kkett {\sigma}\otimes \kkett{\sigma}$ we look for a solution to the equations of motion. Clearly, the solutions that we have found in section \ref{sec:sigma} remain to be solutions and they correspond to  $\kkett {\mathds{1}}\otimes \kkett{\sigma}$, $\kkett {\eps}\otimes \kkett{\sigma}$, $\kkett {\sigma}\otimes \kkett{\mathds{1}}$, and $\kkett {\sigma}\otimes \kkett{\eps}$ boundary states. Ellwood invariants goes in the same fashion as in the case of the single Ising. They are only multiplied by corresponding factors ($\pm$1,0) in accordance with appropriate product with $\kkett{\sigma}$ in the other sector.

\begin{table}[H]
\centering
\begin{tabular}{|r|rrrr|}\hline
Solution                                  & 1        & 2        & 3        & 4        \\   \hline
$\ket B_\Psi$      &$\kkett{\mathds{1}} \otimes \kkett{\mathds{1}}$ & $\kkett{\mathds{1}} \otimes \kkett\eps$ & $\kkett\eps \otimes \kkett {\mathds{1}}$ & $\kkett\eps \otimes \kkett\eps $ \\   \hline
$c_1\ket0$                                & 0.23926  & 0.23926  & 0.23926  &  0.23926  \\
$c_1\ket{\eps ^{(1)}} $               & $-0.16828$ & 0.16828  & $-0.16828$ & 0.16828    \\
$c_1\ket{\eps ^{(2)}} $               & $-0.16828$ & $-0.16828$ & 0.16828  & 0.16828    \\
$c_1\ket{\eps ^{(1)}\eps ^{(2)}}$ & $-0.11836$ & 0.11836  & 0.11836  & $-0.11836$   \\ \hline
\end{tabular}
\caption{Coefficients of the string field for the new solutions on the $\sigma\otimes \sigma$-brane at level one.}
\label{sol-sigma-sigma}
\end{table}

If we look for the action for the field truncated to the level one
\begin{equation}
\ket\Psi =tc_1\ket 0 +ac_1\ket{\eps ^{(1)}}+bc_1\ket{\eps ^{(2)}}+cc_1\ket{\eps ^{(1)}\eps ^{(2)}},
\end{equation}
we can easily find
\begin{eqnarray}
\mathcal{V} (t,a,b,c)  = -\frac{1}{2}t^2-\frac{1}{4}(a^2+b^2)+\frac{1}{3}K^3t^3 +K^2t(a^2+b^2)+2Kabc+Ktc^2.
\end{eqnarray}
From this truncated action the orbifold picture is apparent if we compare it with the action derived in \cite{Karcz} (see also \cite{Bagchi}) for an open string field living on the two parallel D-branes.\footnote{D-branes with separation $1/\sqrt{2}$ correspond to a parameter $d=1$ in their notation. Moreover, substituting $t\rightarrow T_s$, $c\rightarrow X_a$, $a,b\rightarrow \tau \pm \sigma$, setting consistently $\sigma =0$, and dividing whole action by factor of two (not real two D-branes but rather a D-brane with its mirror image are present in our case) we get the same action.}

\begin{figure}
        \centering
        \begin{subfigure}[t]{0.47\textwidth}
                \includegraphics[width=\textwidth]{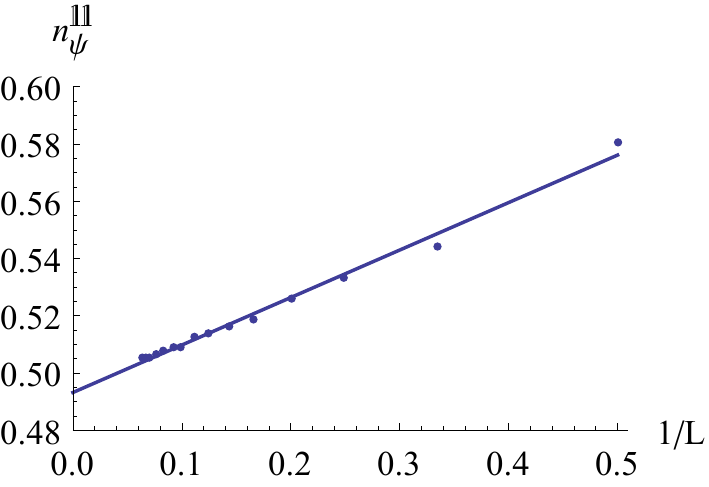}
        \end{subfigure}\qquad
        \begin{subfigure}[t]{0.47\textwidth}
                \includegraphics[width=\textwidth]{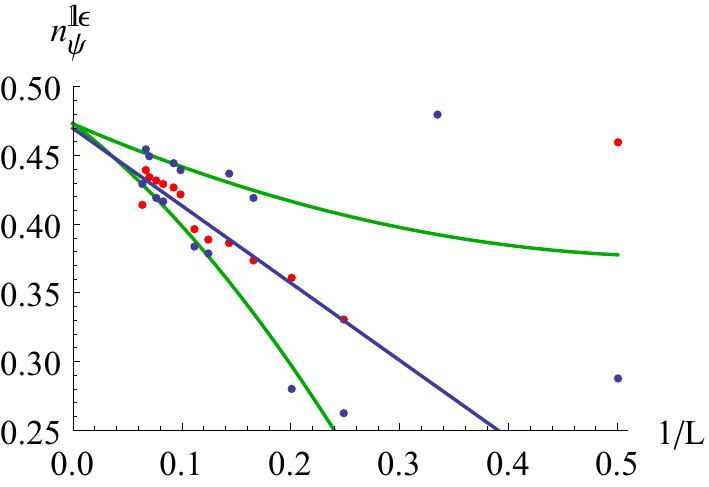}
        \end{subfigure}

        \begin{subfigure}[t]{0.47\textwidth}
                \includegraphics[width=\textwidth]{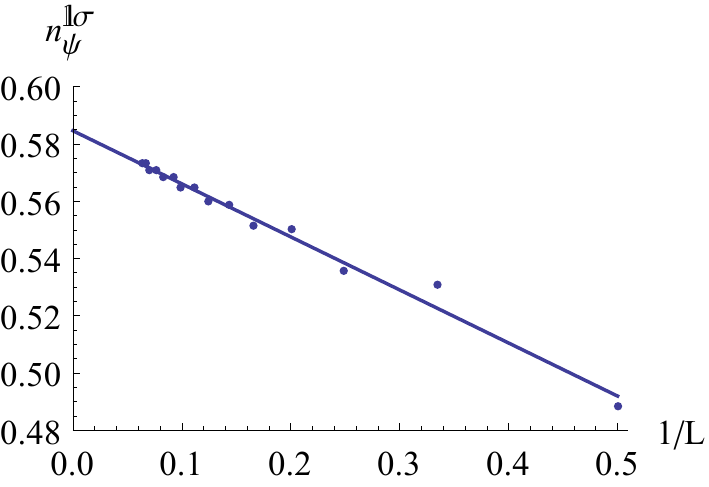}
        \end{subfigure}\qquad
        \begin{subfigure}[t]{0.47\textwidth}
                \includegraphics[width=\textwidth]{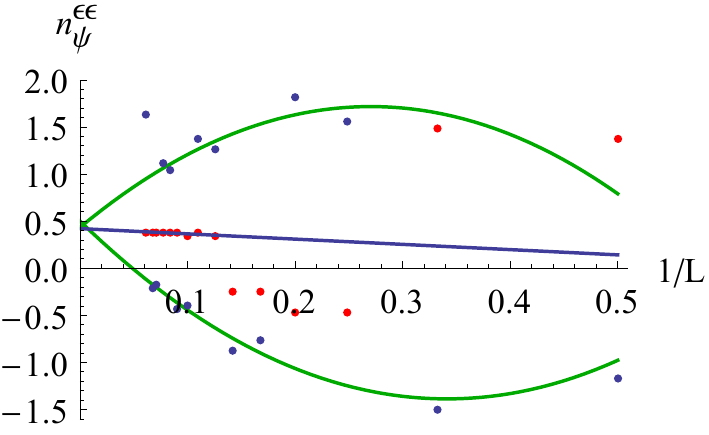}
        \end{subfigure}

        \begin{subfigure}[t]{0.47\textwidth}
                \includegraphics[width=\textwidth]{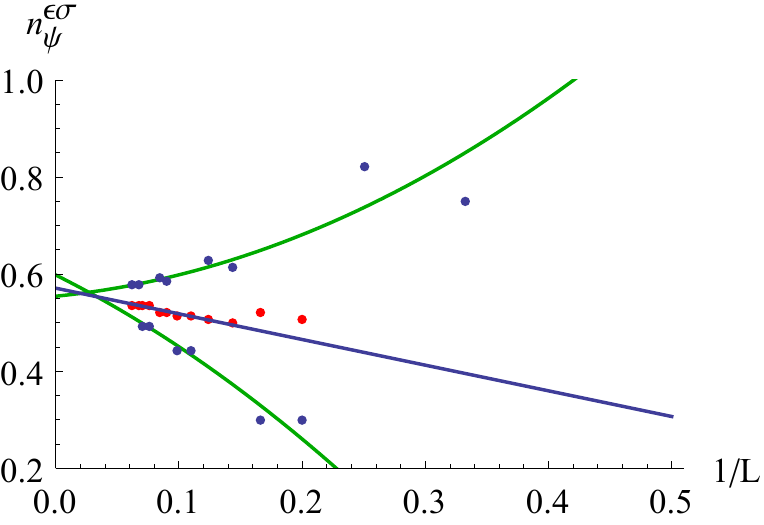}
        \end{subfigure}\qquad
        \begin{subfigure}[t]{0.47\textwidth}
                \includegraphics[width=\textwidth]{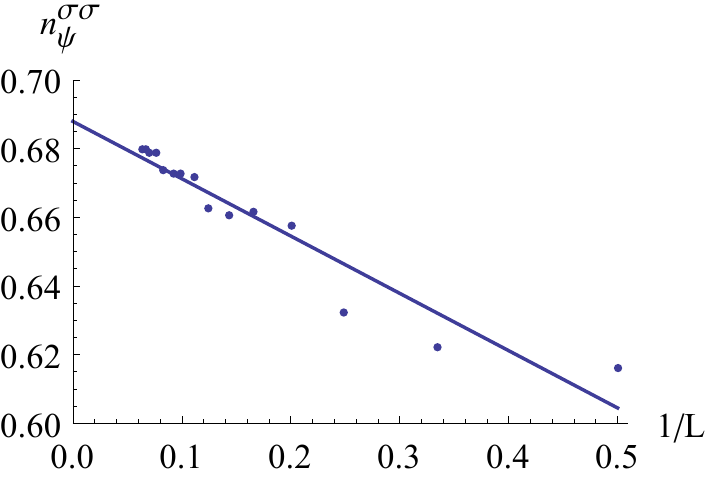}
        \end{subfigure}
        \caption{Gauge invariants for the solution $\Psi$ corresponding to the $\mathds{1}\otimes\mathds{1}$-brane found in the $\sigma\otimes\sigma$-brane background. For all invariants a linear extrapolation to the infinite level is visualized. Due to huge oscillations for some invariants, we also added their Pad\'e-Borel approximation to smoothen the data (red). Moreover, we performed two quadratic fits (green) using the two branches of the oscillating data. Since there are always two nearby points with close values of the invariant, we substitute the two points by one point between them with their mean value in the case of the oscillating data. This substitution enables better extrapolation. We can see quite good agreement of all the fits. We have not been considering the point corresponding to the level $L=16$ for the invariant $n_\Psi^{\eps\eps}$ due to accidental jump that spoils the extrapolation.}\label{double_brane}
\end{figure}

At the level one new four solutions appears. Their coefficients differ only by relative sign and they are mentioned in the table \ref{sol-sigma-sigma}. Picking up one of the solutions and performing the computation up to the level 16 (see table \ref{sol-double ising 2}) we can interpret these solutions as $\kkett{\mathds{1}}\otimes \kkett{\mathds{1}}$,  $\kkett {\mathds{1}}\otimes \kkett{\eps}$, $\kkett {\eps}\otimes \kkett{\mathds{1}}$, and $\kkett{\eps}\otimes \kkett{\eps}$ in double Ising model. Looking at the table \ref{branes} we can interpret this solution in terms of orbifold boson. These solutions correspond to the condensation of the original centered bulk D0-brane that interacts with its mirror image into the fractional D0-brane placed at the orbifold singularity.

\begin{table}[H]
\centering
\begin{tabular}{ |r| l l l l l l l| }\hline
& & & & & & & \\[-2ex]
Level &  Energy    & $ n_\Psi ^{\mathds{1}\mathds{1}}$  &  $n_\Psi ^{\mathds{1}\eps }$    &  $n_\Psi ^{\mathds{1}\sigma }$    &\hspace{0.1mm} $ n_\Psi ^{\eps\eps }$ & \hspace{0.2mm} $n_\Psi ^{\eps\sigma }$ & $ n_\Psi ^{\sigma\sigma }  $             \\   \hline
2        & 0.60317   & 0.58024   & 0.28858   & 0.48785 & -1.15740                & -0.48785                 & 0.61593	 \\
4        & 0.54447   & 0.53344   & 0.26231   & 0.53601 & \hspace{1.5mm}1.54237   & \hspace{0.1mm} 0.81851   & 0.63243	 \\
6        & 0.52870   & 0.51879   & 0.41792   & 0.55168 & -0.75631                & \hspace{0.1mm} 0.29982   & 0.66158	 \\
8        & 0.52138   & 0.51333   & 0.37867   & 0.55949 & \hspace{1.5mm}1.25240   & \hspace{0.1mm} 0.62768   & 0.66279	 \\
10       & 0.51714   & 0.50931   & 0.43869   & 0.56491 & -0.38660                & \hspace{0.1mm} 0.44261   & 0.67291	 \\
12       & 0.51435   & 0.50741   & 0.41674   & 0.56832 & \hspace{1.5mm}1.05267   & \hspace{0.1mm} 0.59137   & 0.67384	 \\
14       & 0.51237   & 0.50559   & 0.44966   & 0.57120 & -0.16680                & \hspace{0.1mm} 0.49358   & 0.67917	 \\
16       & 0.51095   & 0.50544   & 0.42829   & 0.57314 & \hspace{1.5mm}1.64889   & \hspace{0.1mm} 0.57983   & 0.67975	 \\   \hline
$\infty$ & 0.50357   & 0.49316   & 0.49568   & 0.58464 & \hspace{0.1mm} 0.46677                &\hspace{0.1mm} 0.57629                  & 0.68789   \\   \hline
$\infty _{PB}$ & -   & -         & 0.46977   &     -   &\hspace{0.1mm} 0.42077                 &\hspace{0.1mm} 0.57103                  & -         \\   \hline
Expected     & 0.5       & 0.5       & 0.5       & 0.59460 &\hspace{0.2mm} 0.5       & \hspace{0.2mm} 0.59460   & 0.70711	 \\   \hline
\end{tabular}
\caption{Convergence of the boundary coefficients of the new solutions found in the double Ising model. }
\label{sol-double ising 2}
\end{table}

The extrapolated values are obtained using standard linear approximation except of the oscillating coefficients $ n_\Psi ^{\mathds{1}\eps }$, $ n_\Psi ^{\eps\eps }$, and $ n_\Psi ^{\eps\sigma}$. The extrapolation in these cases is the same as in section \ref{sec:sigma}. We have omitted the last value of $ n_\Psi ^{\eps\eps}$ in the extrapolation due to an accidental jump in the coefficient.

Few more solutions with complex coefficients at low levels can be found in the case of doubled Ising model. These solutions may become real as in the case of the solution on the $\mathds{1}$- or $\eps$-brane but it does not happen below the level 16 that we managed to hit.

\section{Discussion and outlook}
\setcounter{equation}{0}

We have demonstrated that numerical approach to OSFT equations of motion yields more solutions and surprises than previously suspected, such as the positive energy solutions describing the $\sigma$-brane of the Ising model formulated on the $\mathds{1}$-brane. An important question is whether by pushing the numerical analysis far enough, with the help of even more powerful computers or by better methods, one can eventually exhaust all physically expected boundary states in the given theory. A question of fundamental importance is whether restricting the computation to Siegel gauge does not prevent us from finding more distant solutions \cite{ET}. Or perhaps, not imposing the gauge choice might allow us to avoid the need to complexify the string field in the intermediate stages of the computation. We have not tried yet to relax this condition, as it would mean significant upgrade of our codes, but we plan to address this in the future. Another issue where one could look for significant improvements is the computation of boundary state coefficients for higher weight primaries. At present, our methods for coefficients of the Ishibashi states with weights $h\gtrsim 1$ give nonconvergent results, and the contributions from different levels have to be resummed by Pad\'e or Pad\'e-Borel methods, which introduces additional source of uncertainty and the results can be used essentially only for qualitative estimates.

Another possible direction is to study the cohomology around these solutions (see \cite{ET_cohom, GI,ES}) from which one should be able to see the spectrum of boundary operators for the new boundary condition and get even more confidence about the physical interpretation. In addition, from the string field theory action for the perturbations one should be able to read off the boundary structure constants of the new theory. Should this all be possible and numerically under control, one could in principle start exploring new boundary conditions in models with $c>1$ where they are not known. This would also be relevant for the study of conformal defects, which via the folding trick can be treated as boundary states in the doubled theory. With the exception of the Ising model, the complete classification of conformal defects is not yet available in any of the minimal models \cite{Runkel}. The string field theory methods could potentially solve the problem, for the time being at least numerically.

Ultimately, the biggest goal of this endeavor is, of course, to find the analytic counterparts for our solutions. The numerical results presented in this work can be viewed as an evidence for their existence and hopefully provide some clues as to how these solutions might look like. Some new ingredients to the standard $K, B, c$ algebra have successfully been used in recent works \cite{KOS,Inatomi,Erler_analytic} and this gives us a hope for more progress in this challenging subject. Constructing generic analytic solutions might lead to lots of detailed information about consistent BCFTs and thus solve the outstanding problem of classifying all the boundary states for a given CFT. Logically, however, it is also possible that these new solutions might be constructible only formally, using the data of the searched-for boundary condition; in that case one would probably not learn much, except perhaps for some new consistency conditions.

String field theory already provides some novel constraints on the form of possible boundary states, see for instance (\ref{boundary states relation}) and it would be nice to learn more about the properties of generic BCFT's on general grounds.

\section*{Acknowledgments}

\noindent

We would like to thank Matthias Gaberdiel, Davide Gaiotto, Ted Erler, Carlo Maccaferri, Masaki Murata, Yuji Okawa, Tom\'{a}\v{s} Proch\'{a}zka, Ingo Runkel, Cornelius Schmidt-Colinet and Barton Zwiebach for useful discussions.

The access to computing and storage facilities owned by parties and
projects contributing to the Czech National Grid Infrastructure
MetaCentrum, provided under the programme "Projects of Large
Infrastructure for Research, Development, and Innovations"
(LM2010005), and to CERIT-SC computing facilities provided under the
programme Center CERIT Scientific Cloud, part of the Operational
Program Research and Development for Innovations, reg. no. CZ.
1.05/3.2.00/08.0144 is highly appreciated.

M.S. would like to thank CERN theory division for the hospitality while this work was in progress, and his stay was also partly supported by the MSMT grant LG13031.
This research was supported by the Grant Agency
of the Czech Republic under the grant P201/12/G028.

\begin{appendix}

\section{Some more solutions on $\mathds{1}$-brane}\label{ap: solutions}
\setcounter{equation}{0}

In this section we present some further solutions we found on $\mathds{1}$-brane using starting points at level~4. We have recurrently improved them and evaluated the corresponding invariants up to level 20.

Unfortunately we do not have a clear and unambiguous interpretation for any of the solutions and we are not even sure whether they are physical. Up to level 20 they still have large imaginary parts and some of them might not converge to real ones. We refrain from showing any extrapolations since due to the slow and oscillatory convergence they depend too sensitively on parameters like number of data points or interpolating function. Even the few relatively stable invariants may not be trustworthy since the behavior of the invariants can change quite dramatically when the solution becomes real as we have seen in section \ref{sec:identity}.

\begin{table}\nonumber
\centering
\footnotesize{
\begin{tabular}{|l|lllll|}\hline
Level  & Energy        & $n_\Psi^\mathds{1}$   & $n_\Psi^\eps$     & $n_\Psi^\sigma$       & $\mbox{Im/Re}$   \\\hline
2  & $-1.17855-2.45419 i$ & $1.55408-0.23558  i$ & $-1.91620+4.83536  i$ & $1.33226+0.47364  i$ & 1.72979 \\
4  & $\hspace{0.28cm} 0.78931-1.78508 i$ & $1.41307-0.05527 i$ & $-2.22469+0.26449 i$ & $0.97012+0.43505 i$ & 1.67605 \\
6  & $\hspace{0.28cm} 1.33064-1.34584  i$ & $1.35690 +0.02281 i$ & $-2.68597+1.40135  i$ & $0.77144+0.47526 i$ & 1.21055 \\
8  & $\hspace{0.28cm} 1.52209-1.03986  i$ & $1.32418+0.05894 i$ & $-2.04892-0.18531 i$ & $0.69125+0.44907 i$ & 0.95045 \\
10 & $\hspace{0.28cm} 1.59184-0.83032 i$ & $1.29648+0.08100  i$ & $-2.33084+0.65181  i$ & $0.61879+0.44087 i$ & 0.80562 \\
12 & $\hspace{0.28cm} 1.61424-0.68266 i$ & $1.27719+0.09299 i$ & $-1.84844-0.24667 i$ & $0.58318+0.41728  i$ & 0.71152 \\
14 & $\hspace{0.28cm} 1.61661-0.57460 i$ & $1.26028+0.10092   i$ & $-2.08891+0.36042  i$ & $0.54486+0.40458 i$ & 0.66427 \\
16 & $\hspace{0.28cm} 1.61012-0.49272 i$ & $1.24772+0.10544  i$ & $-1.72022-0.24690 i$ & $0.52455 +0.38605 i$ & 0.62832 \\
18 & $\hspace{0.28cm} 1.59974-0.42883 i$ & $1.23636+0.10847  i$ & $-1.92949+0.21489 i$ & $0.50056+0.37426 i$ & 0.59972 \\
20 & $\hspace{0.28cm} 1.58779-0.37772 i$ & $1.22751+0.11005  i$ & $-1.63383-0.23515 i$ & $0.48728+0.35948 i$ & 0.57599 \\\hline
\end{tabular}
}
\caption{This first solution can be found already at level 2 and quite likely it might be a combination of two D-branes. The energy could be converging to $\sqrt{2}$, but this seems to be in contradiction with the tendency of the invariants $n_\Psi^\mathds{1}$ which should converge to the same number \cite{Baba}. This strongly reminds us the situation from section \ref{sec:identity} where at some point the solution became exactly real and the $n_\Psi^\mathds{1}$ started converging to the correct value. Without such a change in the behavior, the other invariants would prevent us from matching it to any combination of boundary states (\ref{boundary states}).
}
\end{table}

\begin{table}\nonumber
\centering
\footnotesize{
\begin{tabular}{|l|lllll|}\hline
Level  & Energy        & $n_\Psi^\mathds{1}$   & $n_\Psi^\eps$     & $n_\Psi^\sigma$       & $\mbox{Im/Re}$   \\\hline
4  & $15.8232 -14.9035 i$ & $0.19661-1.35215  i$ & $\hspace{0.28cm} 73.0329 +19.3118  i$ & $-2.13207-0.14212 i$ & 0.63678 \\
6  & $1.87935 -4.39551 i$ & $0.29361-0.68792 i$ & $-5.42787-1.37374  i$ & $-1.65546+0.01740 i$ & 0.61420 \\
8  & $0.81607-3.21150  i$ & $0.38529-0.49078 i$ & $\hspace{0.28cm} 10.5453 +8.33765  i$ & $-1.45902+0.11364 i$ & 0.66266 \\
10 & $0.61653-2.63836 i$ & $0.38971-0.41028  i$ & $-1.30408-0.30582 i$ & $-1.32896+0.13587 i$ & 0.68699 \\
12 & $0.56849-2.27155 i$ & $0.41567-0.35024 i$ & $\hspace{0.28cm} 4.10223 +5.12681  i$ & $-1.25536+0.15252 i$ & 0.71533 \\
14 & $0.55670-2.01096 i$ & $0.41567-0.31296 i$ & $-0.52441+0.11942 i$ & $-1.19805+0.16132  i$ & 0.75775 \\
16 & $0.55456-1.81491 i$ & $0.42670-0.28182 i$ & $\hspace{0.28cm} 1.87039 +3.57380   i$ & $-1.15859+0.16925 i$ & 0.78561 \\
18 & $0.55478-1.66158 i$ & $0.42545 -0.25922 i$ & $-0.37486+0.32254 i$ & $-1.12582+0.17531 i$ & 0.81474 \\
20 & $0.55524-1.53815 i$ & $0.43082-0.23935 i$ & $\hspace{0.28cm} 0.80327+2.66509  i$ & $-1.10112+0.18130	 i$ & 0.84869 \\\hline
\end{tabular}
}
\caption{The second solution is the best match to the $\eps$-brane we have found, but there is a big deviation in energy (which should equal 0.707) and it has also a huge imaginary part which does not seem to decrease uniformly.}
\end{table}

\begin{table}\nonumber
\centering
\footnotesize{
\begin{tabular}{|l|lllll|}\hline
Level  & Energy        & $n_\Psi^\mathds{1}$   & $n_\Psi^\eps$     & $n_\Psi^\sigma$       & $\mbox{Im/Re}$   \\\hline
4  & $5.25280 -29.6447 i$ & $0.18463+0.11448 i$ & $\hspace{0.28cm} 74.5115 -30.4759 i$ & $-2.74394-0.00505 i$ & 0.59302 \\
6  & $4.95530-10.4754 i$ & $0.16760+0.13759 i$ & $-7.07440+13.2339 i$ & $-2.01204+0.02564  i$ & 0.59102 \\
8  & $3.71882-5.83238 i$ & $0.19973+0.15388 i$ & $\hspace{0.28cm} 21.3488 +0.04218 i$ & $-1.72245-0.01537  i $& 0.58886 \\
10 & $2.97511-3.95225 i$ & $0.19958+0.13976 i$ & $-5.16769+6.82781 i$ & $-1.56625-0.00601 i$ & 0.55286 \\
12 & $2.50918-2.97126 i$ & $0.20836+0.13686 i$ & $\hspace{0.28cm} 11.2159 +1.88745 i$ & $-1.46412-0.00978 i$ & 0.53313 \\
14 & $2.19429-2.37778 i$ & $0.20817+0.12908 i$ & $-3.93010 +4.61927 i$ & $-1.39691-0.00817 i$ & 0.53568 \\
16 & $1.96794-1.98283 i$ & $0.21178+0.12561 i$ & $\hspace{0.28cm} 7.24443 +2.09508 i$ & $-1.34565-0.00928 i$ & 0.55190 \\
18 & $1.79734-1.70206 i$ & $0.21128+0.12046 i$ & $-3.23469+3.54717 i$ & $-1.30816-0.00928  i$ & 0.58623 \\
20 & $1.66399-1.49262 i$ & $0.21296+0.11743 i$ & $\hspace{0.28cm} 5.16524 +2.05352 i$ & $-1.27743-0.01012  i$ & 0.61986 \\\hline
\end{tabular}
}
\caption{The third solution. It does not show tendency to become more real at higher levels, although some invariants do. }
\end{table}

\begin{table}\nonumber
\centering
\footnotesize{
\begin{tabular}{|l|lllll|}\hline
Level  & Energy        & $n_\Psi^\mathds{1}$   & $n_\Psi^\eps$     & $n_\Psi^\sigma$       & $\mbox{Im/Re}$   \\\hline
4  & $\hspace{0.28cm} 59.126     -49.7562 i$ & $3.54769+0.05130  i$ & $-55.9217+28.3072 i$ & $\hspace{0.28cm} 0.17951 -0.85771 i$ & 0.83438 \\
6  & $\hspace{0.28cm} 2.86527    -13.7160  i$ & $2.47627+0.13883   i$ & $\hspace{0.28cm} 42.7899 -13.9287 i$ & $-0.59439-1.76930   i$ & 0.79523 \\
8  & $-0.69710  -7.72293 i$ & $2.22963+0.11553   i$ & $-23.3454+12.9107 i$ & $-0.90029-1.08668  i$ & 0.85930 \\
10 & $-0.84556  -4.71321 i$ & $2.09593+0.07366  i$ & $\hspace{0.28cm} 31.8444 +8.26775 i$ & $-0.91008-0.94321 i$ & 0.86416 \\
12 & $-0.56650    -3.11761 i$ & $2.01600  +0.04084  i$ & $-5.70906+4.08035 i$ & $-0.82865-0.74664 i$ & 0.86321 \\
14 & $-0.26486  -2.19043 i$ & $1.95414+0.02035  i$ & $\hspace{0.28cm} 20.4005 +4.90175 i$ & $-0.77314-0.66680 i$ & 0.85356 \\
16 & $-0.00526-1.60929 i$ & $1.91151+0.00447 i$ & $-1.48657+2.49446 i$ & $-0.71261-0.56743 i$ & 0.87044 \\
18 & $\hspace{0.28cm} 0.20770   -1.22308 i$ & $1.87484-0.00581 i$ & $\hspace{0.28cm} 15.4965 +3.23866 i$ & $-0.67058-0.51903 i$ & 0.89005 \\
20 & $\hspace{0.28cm} 0.38108   -0.95440i$ & $1.84740 -0.01414  i$ & $\hspace{0.28cm} 0.15867 +1.65249 i$ & $-0.62722-0.45963  i$ & 0.90270 \\\hline
\end{tabular}
}
\caption{The fourth solution. It does not show tendency to become more real at higher levels, although some invariants do.}
\end{table}

\begin{table}\nonumber
\centering
\footnotesize{
\begin{tabular}{|l|lllll|}\hline
Level  & Energy        & $n_\Psi^\mathds{1}$   & $n_\Psi^\eps$     & $n_\Psi^\sigma$       & $\mbox{Im/Re}$   \\\hline
4  & $-88.6270 -841.048 i$ & $4.03869 -2.99405  i$ & $-349.429 -171.787 i$ & $-1.08430-11.8809   i$ & 0.90573 \\
6  & $\hspace{0.28cm} 5.97346 -6.18772 i$ & $0.71804-0.37139 i$ & $\hspace{0.28cm} 15.8499  +78.5997 i$ & $\hspace{0.28cm} 1.35640 -0.80274  i$ & 0.97248 \\
8  & $\hspace{0.28cm} 7.87409 -3.83051 i$ & $0.60002-0.23520 i$ & $-16.4228 +23.6466 i$ & $-0.12968+0.62655 i$ & 0.86424 \\
10 & $\hspace{0.28cm} 4.15567 -3.31309 i$ & $0.59799 -0.20503 i$ & $\hspace{0.28cm} 3.23982  +13.8598 i$ & $-0.30361+0.41484 i$ & 0.87972 \\
12 & $\hspace{0.28cm} 2.63239 -2.91809 i$ & $0.59721-0.17018 i$ & $-9.58180  +11.2610  i$ & $-0.37240+0.45549 i$ & 0.89492 \\
14 & $\hspace{0.28cm} 1.85777 -2.60619 i$ & $0.58392-0.16094 i$ & $\hspace{0.28cm} 0.25318 +8.38227 i$ & $-0.43405+0.36492 i$ & 0.93905 \\
16 & $\hspace{0.28cm} 1.40997 -2.35546 i$ & $0.58101 -0.14422 i$ & $-6.77543 +7.68336 i$ & $-0.45303 +0.35651 i$ & 0.93655 \\
18 & $\hspace{0.28cm} 1.12734 -2.15012 i$ & $0.57116-0.13912 i$ & $-0.72274+6.24676 i$ & $-0.48514+0.30211 i$ & 0.86293 \\
20 & $\hspace{0.28cm} 0.93725-1.97894 i$ & $0.56936-0.12913  i$ & $-5.29955 +6.15556 i$ & $-0.49300+0.28408 i$ & 0.82501 \\\hline
\end{tabular}
}
\caption{This last solution shows a bit more tendency to become real, however, due to the still very significant imaginary part it is perhaps pointless to offer any interpretation. One important caveat about this solution is that the starting point at level 4 did not lead to any solution at level 6 with prescribed precision. Nevertheless, the final point of the iteration (fairly random)  was used as the new starting point at level 8 and from that point on it did converge.}
\end{table}


\section{Closer look at the double Ising}\label{ap: double_ising}
\setcounter{equation}{0}

This appendix contains notes on the correspondence of the double Ising model and free-boson living on the orbifold. The spectrum of bulk primary operators have been found as a decoupling point of Ashkin-Teller model, i.e. model consisting of two Ising model slices coupled together by a quartic term. Precise correspondence between primary fields in the Ising model picture and the free boson picture (up to sign ambiguities) can be fixed considering OPE's. There is a ambiguity in fixing the sign in the correspondence of $\cos\left (\frac{X}{\sqrt{2}}\right )$ as well as $\cos\left (\sqrt{2}\tilde{X}\right )$ that cannot be fixed using OPE's of the mentioned fields.

Every spinless primary from the table \ref{primaries} has corresponding Ishibashi state that will be denoted as
\[  \begin{array}{l l l}
   \kett{n^2},&  \kett{\frac{(n+1)^2}{2},1},& \kett{\frac{(n+1)^2}{2},2},\\
    \kett{\frac{(2n+1)^2}{8}},&  \kett{\frac{(2n+1)^2}{16},1},&  \kett{\frac{(2n+1)^2}{16},2},
  \end{array}\]
where we have to take the multiplicity of primary fields into account. The states with the multiplicity two correspond to the states that switches under the switch of the two Ising models Ising$_1\leftrightarrow$ Ising$_2$. We have for example correspondence
\be
\kett{0}\otimes \kett{\eps} = \sum _{n=0}^{\infty}\kett{\frac{(2n+1)^2}{2},2},\qquad \kett{\eps}\otimes \kett{0} = \sum _{n=0}^{\infty}\kett{\frac{(2n+1)^2}{2},1}
\ee
and similarly for all the other states.

We have already reviewed boundary states in table \ref{branes}. Tensor products of boundary states of the single Ising model are mentioned explicitly. All these boundary states can be interpreted as D-branes of the orbifolded boson. One can use corresponding one-point function on the disk and correspondence between bulk primaries from the table \ref{primaries} to clarify the correspondence. An energy of D-brane can be computed by $\langle \mathds{1}\rangle$. Coefficient $\frac{\langle \partial X \bar{\partial}X\rangle }{\langle \mathds{1}\rangle}$ characterizes a nature of the boundary condition (following from the propagator on the disk we get +1 in the case of Dirichlet boundary condition and -1 in the case of Neumann boundary condition). These two coefficients gives us interpretation from the second column of the table. From the $\langle \cos (\frac{X}{\sqrt{2}})\rangle$ coefficient one finds a position of D-branes. In the case of D1 branes do not have interpretation of D-brane position and $\cos (\sqrt{2}\tilde{X})$ corresponding to the position on the T-dual circle alternates this field.

There are two continuous families of boundary states denoted by $\kkett{D(\phi)}$ and $\kkett{N(\tilde{\phi})}$. The first family corresponds to marginal deformations of the centered bulk D0-brane by the marginal operator $\eps \otimes \eps$ or equivalently $\partial X \bar{\partial}X$ in the orbifold picture corresponding to moving the original D-brane.The second one consists of the T-dual D-branes.

In the free boson picture the two continuous families correspond to the orbifold projections generic boundary states of free boson living on the circle $\kkett{D(\phi)}$ and $\kkett{N(\tilde{\phi})}$, namely
\be
\kkett{D(\phi)_O}=\frac{1}{\sqrt{2}}[\kkett{D(\phi)}-\kkett{D(-\phi)}], \qquad \kkett{N(\tilde{\phi})_O}=\frac{1}{\sqrt{2}}[\kkett{N(\tilde{\phi})}-\kkett{N(-\tilde{\phi})}]
\ee
for $\phi \in (0,\pi R)$ and $\tilde{\phi} \in (0,\pi /R)$. We can also express boundary states from the two continuous families as combinations of the Ashkin-Teller Ishibashi states according to \cite{Oshikawa}. To write desired boundary states in a simple form let us consider symmetric and antisymmetric combinations of the Ashkin-Teller Ishibashi states
\begin{eqnarray}\nonumber
  \kett{\frac{n^2}{8},S}&=&\left\{
  \begin{array}{l l}
    \frac{1}{\sqrt{2}}[\kett{\frac{(2k)^2}{8},1}+\kett{\frac{(2k)^2}{8},2}] & \quad \text{if $n=2k$}\\
    \kett{\frac{(2k+1)^2}{8}} & \quad \text{if $n=2k+1$}
  \end{array} \right . ,\\
  \kett{\frac{n^2}{2},A}&=&\frac{1}{\sqrt{2}}[\kett{\frac{n^2}{2},1}-\kett{\frac{n^2}{2},2}].
  \end{eqnarray}
Note that these states are symmetric (antisymmetric) with respect to the switch Ising$_1\leftrightarrow$ Ising$_2$. Now, we are ready to write explicit expression for the two families as
\begin{eqnarray}\nonumber
\kkett{D_O(\phi)}&=&\sum _{k=0}^{\infty}\kett{n^2}+\sqrt{2}\sum _{k=1}^{\infty}\cos\left (\frac{k\phi}{\sqrt{2}}\right )\kett{\frac{k^2}{8},S}\\
\kkett{N_O(\tilde{\phi})}&=&\sqrt{2}\sum _{k=0}^{\infty}\kett{n^2}+2\sum _{k=1}^{\infty}\cos \left (\sqrt{2}k\tilde{\phi}\right )\kett{\frac{k^2}{2},A}
\end{eqnarray}
for $\phi \in (0,\pi R)$ and $\tilde{\phi} \in (0,\pi /R)$. Note that the energy of states from the first family is indeed 1, whereas the energy of the states from the second one is $\sqrt{2}$. From the overlap with $\ket{\frac{1}{8},S}$, corresponding to $\pm \sqrt{2}\cos \left (\frac{X}{\sqrt{2}} \right )$, we can see that position of the $\kkett{D_O(\phi)}$ is indeed $R\phi$. Similarly, from the overlap with $\ket{\frac{1}{2},A}$ corresponding to $\pm \sqrt{2}\cos (\sqrt{2}\tilde{X})$ we can see that $\tilde{\phi}/R$ is the position of the D-brane in the T-dual picture.

Until now we have been discussing the the continuous families corresponding to the orbifold projection of boundary states for the free boson on the circle. When considering the free boson on the orbifold a new sector of so called twisted primary states appears and one can suspect that new primary fields emerge. There exist a discrete set of eight more boundary states corresponding to fractional D-branes
\begin{eqnarray}\nonumber
\kkett{D_{O}(\phi _0)\pm}=\frac{1}{\sqrt{2}}\kkett{D(\phi _0)}\pm \frac{1}{\sqrt [4]{2}}\ket{D(\phi _0)_T},\\
\kkett{N_{O}(\tilde{\phi} _0)\pm}=\frac{1}{\sqrt{2}}\kkett{N(\phi _0)}\pm \frac{1}{\sqrt [4]{2}}\ket{N(\phi _0)_T},
\end{eqnarray}
where $\phi _0$ takes values $0,\pi R$ and $\tilde{\phi} _0$ takes values $0,\pi /R$. The first parts $\kkett{D(\phi _0)}$ and $\kkett{N(\phi _0)}$ correspond to the contribution from the untwisted sector and we denoted in the twisted sector
\begin{eqnarray}\nonumber
\ket{D(0)_T}&=&\frac{1}{\sqrt{2}}(\kett{\sigma\otimes\mathds{1}}+\kett{\sigma\otimes\eps}+\kett{\mathds{1}\otimes\sigma}+\kett{\eps\otimes\sigma})\\ \nonumber
\ket{D(\sqrt{2}\pi)_T}&=&\frac{1}{\sqrt{2}}(\kett{\sigma\otimes\mathds{1}}+\kett{\sigma\otimes\eps}-\kett{\mathds{1}\otimes\sigma}-\kett{\eps\otimes\sigma})\\ \nonumber
\ket{N(0)_T}&=&\kett{\sigma\otimes\mathds{1}}-\kett{\sigma\otimes\eps}\\
\ket{N(\pi /\sqrt{2})_T}&=&\kett{\mathds{1}\otimes\sigma}-\kett{\eps\otimes\sigma}.
\label{fractional_branes}
\end{eqnarray}
These fractional D-branes can be identified with tensored boundary states in the Ising model picture. Note that fractional D-branes have half of the energy of corresponding bulk D-brane and they are located at the orbifold singularity.

Characteristics of D-branes discussed so far do not allow precise identification of a D-brane since some fractional D-branes share the same values. We have to introduce some twist charge to distinguish them. We will call the sign in the $\kkett{D_{O}(\phi _0)\pm}$ and $\kkett{N_{O}(\tilde{\phi} _0)\pm}$ the desired twist charge. If we cross out all the descendants in (\ref{fractional_branes}) to avoid non-normalizable states, the four states form orthogonal set. The twist charge is then computable as an an overlap of the boundary state with the sum of these four states divided by 2.

The correspondence between the primaries in the BCFT is a bit more complicated. In the following, show examples up to level 2.

$\mathds{1}\ot \mathds{1}$ boundary condition for the Ising model corresponds to the fractional D0-brane at $X=0$. In the Ising sector we have only primaries constructed over identity, on the orbifold side we have identity, $\partial X$ and winding modes. However $\partial X$ and half of the winding states are removed by the orbifold projection. The surviving states are

\begin{equation}\nonumber
\begin{array}{|c|c|c|}\hline
h  & (\rm{Ising})^2                 & \rm{Orbifold}\\\hline
0  & \mathds{1}                     & \mathds{1}\\
2  & T\ot\mathds{1}-\mathds{1}\ot T & -\cos \sqrt{2}\tilde X\\\hline
\end{array}
\end{equation}
To fix signs in the correspondence of boundary operators one has to consider also bulk-boundary correlators.

$\mathds{1}\ot \sigma$ boundary condition for the Ising corresponds to the fractional D1-brane. We have additional $\mathds{1}\ot \eps$ primary in the Ising sector. On the D1-brane momentum modes instead of winding modes are present. The orbifold projection once again removes $\partial X$ and half of the momentum modes

\begin{equation}\nonumber
\begin{array}{|c|c|c|}\hline
h            & (\rm{Ising})^2 & \rm{Orbifold}\\\hline
0            & \mathds{1}                     & \mathds{1}\\
\frac{1}{2}  & \mathds{1}\ot \eps         & \pm\sqrt{2}\cos \frac{X}{\sqrt{2}}\\
2            & T\ot\mathds{1}-\mathds{1}\ot T & -\cos \sqrt{2} X\\\hline
\end{array}
\end{equation}
The unfixed sign above is the same as in the correspondence for $\sigma\ot\sigma$ in the bulk.

The situation for $\sigma\ot \sigma$ boundary condition, which corresponds to the bulk D0-brane located at $X=\frac{\pi}{\sqrt{2}}$, is the most complicated. The primaries in the Ising sector are constructed over all four combinations of $\mathds{1}$ and $\eps$. The free boson sector can be described as strings on two D0-branes with Chan-Paton-like description, where the spectrum is reduced by the orbifold projection. The $Z_2$ symmetry acts on the 2x2 matrix as
\be
Z\left(\begin{array}{cc} a & b \\ c & d \\ \end{array}\right)=\left(\begin{array}{cc} d & c \\ b & a \\ \end{array}\right).
\ee
The primary states in the normal sector are the identity, $\partial X$ and winding modes with integer winding number, all multiplied the Chan-Paton factors. Unlike the previous cases now we have a nontrivial twisted sector (that appears as the off-diagonal components of the matrices) that describes states going from the D0-brane to its mirror image. The primaries in the twisted sector are winding modes with half integer winding times the Chan-Paton factors. Up to the signs, which we will explain later, the match between the primary states is

\begin{equation}\nonumber
\begin{array}{|c|c|c|}\hline
h           & (\rm{Ising})^2                                             & \rm{Orbifold}\\\hline
0           & \mathds{1}\ot \mathds{1}                                   & \left(\begin{array}{cc} 1 & 0 \\ 0 & 1 \\ \end{array}\right)\\
\frac{1}{2} & \mathds{1}\ot \eps                                     & \pm\sqrt{2}\cos \frac{\tilde X}{\sqrt{2}} \left(\begin{array}{cc} 0 & 1 \\ 1 & 0 \\ \end{array}\right)\\
\frac{1}{2} & \eps\ot \mathds{1}                                     & \pm\sqrt{2}\sin \frac{\tilde X}{\sqrt{2}} \left(\begin{array}{cc} 0 & i \\ -i & 0 \\ \end{array}\right)\\
1           & \eps\ot \eps                                       & \pm i \sqrt{2}\partial X \left(\begin{array}{cc} 1 & 0 \\ 0 & -1 \\ \end{array}\right) \\
2           & \partial \eps\ot \eps-\eps\ot \partial\eps & \pm 2i\sin \sqrt{2}\tilde X  \left(\begin{array}{cc} 1 & 0 \\ 0 & -1 \\ \end{array}\right) \\
2           & T\ot \mathds{1}-\mathds{1}\ot T                            & \pm \cos \sqrt{2}\tilde X  \left(\begin{array}{cc} 1 & 0 \\ 0 & 1 \\ \end{array}\right)\\\hline
\end{array}
\end{equation}

From the OPE alone we cannot determine the two emphasized signs at level $1/2$ and we can also exchange these two states leaving 8 possible correspondences. Once we fix these ambiguities, all signs at higher levels are uniquely determined. The undetermined quantities are fixed by the bulk-boundary correlators $\la \sigma\ot \sigma (z,\bar z) \eps\ot \eps (x) \ra$, $\la \mathds{1}\ot \sigma (z,\bar z) \mathds{1}\ot \eps (x) \ra$ and $\la \sigma\ot \mathds{1} (z,\bar z) \eps\ot \mathds{1} (x) \ra$. The first correlator can be easily computed, but we cannot evaluate the other two in the orbifold picture, because we do not know the exact form of the twisted vertex operators. We expect an ambiguity in the bulk correspondence in the twisted sector anyway, so the best that can be possibly done is to relate the bulk and boundary ambiguities.

\end{appendix}

\end{document}